\documentclass[fleqn,usenatbib]{mnras}
\usepackage{newtxtext,newtxmath}
\usepackage[T1]{fontenc}
\usepackage{ae,aecompl}

\usepackage{graphicx}	
\usepackage{amsmath}
\usepackage{amssymb}
\usepackage{url}

\usepackage[normalem]{ulem}

\title[X-ray plateaus from off-axis GRBs]{X-ray Plateaus in Gamma Ray Bursts' light-curves from jets viewed slightly off-axis}

\author[Beniamini et al.]{
Paz Beniamini$^{1,3}$\thanks{E-mail: paz.beniamini@gmail.com},
Rapha\"{e}l Duque$^{2}$,
Fr\'{e}d\'{e}ric Daigne$^{2}$
and Robert Mochkovitch$^{2}$
\\
$^{1}$Department of Physics, The George Washington University, Washington, DC 20052, USA\\
$^{2}$Sorbonne Universit\'{e}, CNRS, UMR 7095, Institut d'Astrophysique de Paris, 98 bis boulevard Arago, 75014 Paris, France\\
$^{3}$Division of Physics, Mathematics and Astronomy, California Institute of Technology, Pasadena, CA 91125, USA}

\date{Accepted XXX. Received YYY; in original form ZZZ}

\pubyear{2020}

\begin{document}
\label{firstpage}
\pagerange{\pageref{firstpage}--\pageref{lastpage}}
\maketitle

\begin{abstract}
Using multiple observational arguments, recent work has shown that cosmological GRBs are typically viewed at angles within, or close to the cores of their relativistic jets. One of those arguments relied on the lack of tens-of-days-long periods of very shallow evolution that would be seen in the afterglow light-curves of GRBs viewed at large angles. Motivated by these results, we consider that GRBs efficiently produce $\gamma$-rays only within a narrow region around the core. We show that, on these near-core lines-of-sight, structured jets naturally produce shallow phases in the X-ray afterglow of GRBs. These plateaus would be seen by a large fraction of observers and would last between $10^2-10^5$ s. They naturally reproduce the observed distributions of time-scales and luminosities as well as the inter-correlations between plateau duration, plateau luminosity and prompt $\gamma$-ray energy. An advantage of this interpretation is that it involves no late time energy injection which would be both challenging from the point of view of the central engine and, as we show here, less natural given the observed correlations between plateau and prompt properties.
\end{abstract}



\section{Introduction}
\label{sec:Intro}
The angular structure of Gamma Ray Burst (GRB) jets is a topic of major importance both for the prospect of obser\-ving GRB signals from upcoming gravitational wave detections of binary neutron star mergers, and for enhancing our understanding of the formation and propagation of ultra-relativistic jets \citep{Lamb2017,Kathirgamaraju2018,Granot2018,Beniamini2019,Gill2019,Beniamini2020}.
As we show in this work, it may also have other observational signatures, namely in the afterglow light-curves.

\cite{BN2019} have recently considered the ratio between the energy emitted in the early X-ray afterglow of long GRBs to the energy emitted in $\gamma$-rays during their prompt phase. They have shown that, since observationally this ratio does not vary strongly between different bursts, models that do not exhibit a very steep drop of the kinetic energy beyond the core must produce $\gamma$-rays efficiently only up to an angle of $\theta_{\gamma}\lesssim 2\theta_j$, where $\theta_j$ is the opening angle of the core. This conclusion is further supported by two additional independent observations. The first is the relatively small fraction of bursts with low to moderate luminosities, and the second is the lack of X-ray light-curves that evolve as $t^{-0.5}$ or shallower for extended (days to tens of days) periods of time.

The latter observation is of particular interest to us in the present work. Although GRB afterglows do not exhibit tens-of-days-long phases of shallow evolution, a significant fraction of GRBs do in fact exhibit plateaus lasting from hundreds to tens of thousands of seconds after the burst \citep{Nousek2006}. This kind of behaviour is an expected consequence of the forward shock emission produced by structured jets viewed off-axis, provided that, as argued by \cite{BN2019}, bursts are always detected at angles not much greater than $\theta_j$. Indeed early on after the discovery of X-ray plateaus, \cite{Eichler2006} suggested that  plateaus could be the result of structured jets viewed at latitudes beyond the jets' cores. More recently, \cite{Oganesyan2019} have suggested a related but distinct interpretation, in which plateaus are the result of the {\it prompt emission} photons produced at a large angular distance from the observer (i.e., produced by material moving at an angle more than $1/\Gamma$ from the line of sight) and received by the observer at later times.

Many other interpretations of plateaus have been suggested in the literature over the years. Some examples are late-time energy injection from the central engine, shining through an internal process \citep{Ghisellini2007,BM2017} or as the fresh material joins the external shock \citep{Nousek2006,Zhang2006}, forward shock emission from an inhomogeneous jet \citep{Toma2006} (which is a superposition of off-axis emitting regions), 
forward shock emission with time-dependent microphysical parameters \citep{Granot2006,Ioka2006,Panaitescu2006}, contributions from reverse shock emission \citep{Uhm2007,Genet2007,Hascoet2014}, external shock emission in the thick-shell regime \citep{Leventis2014} and delayed afterglow deceleration \citep{GK2006,Kobayashi2007,Shen2012,Duffell2015}.

Since the initial discovery of X-ray plateaus by {\it Swift}, many more plateaus have been observed and their statistics and correlations with other burst properties studied in detail \citep{D08, Margutti2013,D17,Tang2019}. We show that the forward shock emission of GRBs viewed beyond their jet cores can naturally account for these observed correlations without any need to invoke late time energy injection, which is challenging from the point of view of the central engine and, as we show here, less natural given the observed correlations.
Although some plateaus end with a very rapid temporal decline that is clearly inconsistent with an external shock origin \citep{Zhang2006,Liang2006,Troja2007,BM2017}, there are less than a handful of such cases. The vast majority of plateaus are compatible with the geometric or dynamical interpretations we adopt here.
Furthermore, the fraction of bursts with plateaus puts strong constraints on the region within which prompt $\gamma$-rays are efficiently produced (consistent with the results by \citealt{BN2019}) and their typical durations restrict the allowed structure of energy and Lorentz factor beyond the jets' cores. 

The paper is organized as follows. In \S~\ref{sec:modeling} we outline the basic model considered in this work for the calculation of the prompt, afterglow and early steep decline phases in structured jets. We turn in \S~\ref{sec:Plateauoffcore} to describe two classes of plateaus that can be viewed from lines-of-sight which are slightly beyond the jet cores. We obtain the light-curves corresponding to both cases and compare the correlations between the observed parameters with GRB data. We then discuss some general implications of the interpretation presented in this paper in \S~\ref{sec:Discuss} and finally conclude in \S~\ref{sec:conclusion}.

\section{Observed gamma-rays and X-rays from structured jets}
\label{sec:modeling}
Consider a jet in which the kinetic energy per unit solid angle as measured in the central engine frame, $\epsilon$, and the initial Lorentz factor, $\Gamma_0$ of the material can depend on the polar angle from the jet's axis, $\theta$, but (assuming azimuthal symmetry) not on $\phi$. For the purpose of deriving analytic expressions we focus in this paper on power-law profiles for $\epsilon, \Gamma_0$:
\begin{equation}
\epsilon(\theta)=\frac{dE}{d\Omega}=\epsilon_j\left\{ \begin{array}{ll}1 & \theta<\theta_j\ ,\\
\bigg(\frac{\theta}{\theta_j}\bigg)^{-\alpha} & \theta \geq \theta_j\ ,
\end{array} \right.
\end{equation}

\begin{equation}
\label{eq:GAPL}
\Gamma_0(\theta)=1+(\Gamma_j-1)\left\{ \begin{array}{ll}1 & \theta<\theta_j\ ,\\
\bigg(\frac{\theta}{\theta_j}\bigg)^{-\beta} & \theta \geq \theta_j\ ,
\end{array} \right.
\end{equation}
where $\theta_j$ is the opening angle of the jet's core, and $\epsilon_j,\Gamma_j$ are the kinetic energy per solid angle and initial Lorentz factor at the core. 
Part of the energy described by $\epsilon(\theta)$ powers the $\gamma$-rays (prompt emission), while the rest remains as kinetic energy for the blast-wave during the afterglow phase. Denoting by $\eta_{\gamma}$ the $\gamma$-ray efficiency, the post-prompt isotropic-equivalent kinetic energy of the blast wave is $E_{\rm k}(\theta)=4\pi (1-\eta_{\gamma}) \epsilon(\theta)$ and the isotropic-equivalent contribution to the emitted $\gamma$-ray energy along $\theta$ is $E_{\gamma,\rm em}(\theta)=\eta_{\gamma}4\pi \epsilon(\theta)$.
For clarity, and in order to avoid adding further unknown free parameters to the model, we assume in the following that $\eta_{\gamma}$ is constant up to some polar angle $\theta_{\rm max}$ and then vanishes. This is consistent with the aforementioned result on the $\gamma$-efficiency being limited to a certain angle $\theta_\mathrm{max}\la 2\theta_\mathrm{j}$ around the core \citep{BN2019}.

The observed isotropic-equivalent $\gamma$-ray energy at a viewing angle $\theta_{\rm v}$ is then given by
\begin{equation}
\label{eq:gammaenergy}
E_{\gamma}(\theta_{\rm v})=\eta_{\gamma}\int^{\theta_{\rm max}} \frac{\epsilon(\theta)}{\Gamma_0(\theta)^4 (1-\beta_0(\theta) \cos \chi)^{3}}  d\Omega \, ,
\end{equation}
where $\chi$ is the angle between the line of sight and the direction of motion of the emitting material, given by
\begin{equation}
\cos \chi = \cos \theta_{\rm v} \cos \theta + \sin \theta_{\rm v} \sin \theta \cos \phi
\end{equation}
Defining $\Delta \theta \equiv \theta_{\rm v}-\theta_j$ and $q \equiv |\Delta \theta|\Gamma_j$, performing the integration in Eq.~\ref{eq:gammaenergy} allows one to derive a useful approximation to the observed isotropic-equivalent $\gamma$-ray energy \citep{Kasliwal2017,Granot2018,Ioka2018}:
\begin{eqnarray}
\label{eq:Egamma}
& E_{\gamma}(\theta_{\rm v})= \eta_{\gamma}4\pi \epsilon_j
\\ & \times \left\{ \begin{array}{ll}1  \! & 0<\theta_{\rm v}< \theta_j \, ,\\
\max[\frac{\epsilon(\theta_{\rm v})}{\epsilon_j}\Theta(\theta_{\rm max}\!-\!\theta_{\rm v}),(1 + q^{2})^{-2}] \! \! & \! \theta_j<\theta_{\rm v}<2\theta_j \, ,\\ \max[\frac{\epsilon(\theta_{\rm v})}{\epsilon_j}\Theta(\theta_{\rm max}\!-\!\theta_{\rm v}),q^{-2}(1 + q^{2})^{-2}(\theta_j\Gamma_j)^2]\!\! &\! 2\theta_{j} < \theta_{\rm v} \, ,
\end{array} \right.  \nonumber
\end{eqnarray}
where $\Theta(x)$ is the Heaviside function.
\footnote{{Eq.~\ref{eq:Egamma} is given here for the sake of clarity and qualitative understanding. The quantitative calculation performed in this paper are done by directly performing the integral in Eq.~\ref{eq:gammaenergy} and do not involve these approximations.}}. 
So long as the ratio $\frac{\epsilon(\theta_{\rm v})}{\epsilon_j}$ in these expressions dominates, the result is that the observed $\gamma$-ray energy equals the emitted energy along the line of sight: $E_{\gamma}(\theta_{\rm v})=E_{\gamma,\rm em}(\theta_{\rm v})$. This is typically the case for angular energy profiles that are not extremely steep and for viewing angles that are moderate (e.g., $\theta_{\rm v}\lesssim 2\theta_j$ as considered in this paper; see Fig.~2 of 
\citealt{BN2019})

As the blast wave pushes into the external medium it starts decelerating. Seen from an angle $\theta_{\rm v}$, the Lorentz factor of material at position $\theta, \phi$ remains roughly constant up to an observer time (omitting cosmological redshift corrections):
\begin{eqnarray}
\label{eq:tdec}
& \! t^{\theta_{\rm v}}_{\rm dec}(\theta, \phi)\!=\!(1\!-\!\beta_0 \cos \chi)\times
\left\{ \begin{array}{ll}\! \bigg(\frac{17 E_{\rm k}}{8 \pi n m_p c^5}\bigg)^{1/3}\beta_0^{-5/3}\Gamma_0^{-2/3}  & \! \! \mbox{ ISM},\\ \\
\!\frac{9 E_{\rm k}}{16\pi A \beta_0^3 c^3 \Gamma_0^2} \! \! & \! \! \mbox{ wind},
\end{array} \right.
\end{eqnarray}
where $E_{\rm k},\beta_0,\Gamma_0$ are evaluated at $\theta$ and we have assumed an ultra-relativistic blast-wave to set the numerical normalization. The first line holds for a uniform external medium (ISM) of density $n$, and the second line holds for a stellar wind external medium, in which the density varies as $n=(A/m_p)r^{-2}$, where we refer to $A$ as the `wind parameter'. We define, as usual the dimensionless quantity $A_*\equiv A/(5\times 10^{11}\mbox{ g cm}^{-1})$. For material moving along the line of sight, $(1-\beta_0 \cos \chi)\propto \Gamma_0^{-2}$ and one retains the well known relations for the deceleration time, i.e., $t_{\rm dec}\propto \Gamma_0^{-8/3}$ for an ISM and $t_{\rm dec}\propto \Gamma_0^{-4}$ for a wind. We denote these special cases as $t_{\rm d,los} = t^{\theta}_{\rm dec}(\theta, \phi = 0)$.

After $t^{\theta_{\rm v}}_{\rm dec}$, the Lorentz factor starts decreasing. Its value for material at position $\theta, \phi$ at a later time $t > t^{\theta_{\rm v}}_{\rm dec}(\theta, \phi)$ is such that Eq.~\ref{eq:tdec} still holds when replacing $t^{\theta_{\rm v}}_{\rm dec}$ by $t$, hence providing an implicit equation for the Lorentz factor of any portion of the jet at observer time $t$. For material moving along the line of sight (henceforth denoted as `line-of-sight material') this yields simply $\Gamma=\Gamma_0(t/t_{\rm d,los})^{-3/8}$ for an ISM environment and $\Gamma=\Gamma_0(t/t_{\rm d,los})^{-1/4}$ for a wind.

Using the implicit Eq.~\ref{eq:tdec} to determine the evolution of the Lorentz factor with observer time, it is then straightforward to calculate the afterglow luminosity, given a description of the angle-dependent shock-frame emitted luminosity, from a similar integration to that described for the $\gamma$-rays above, in Eq.~\ref{eq:gammaenergy}. The shock-frame emitted luminosity is calculated using the standard forward shock synchrotron radiation (see, e.g., \citealt{WG1999,GS2002}) with corrections to the electron cooling due to synchrotron self Compton (SSC, see \citealt{Beniamini2015} for details).

As it decelerates, energy density gradients within the jet will lead to lateral spreading. For the near-core lines of sight we consider here, this spreading may affect afterglow predictions, especially when considering steep jet structures. 
In Appendix~\ref{sec:spread}, we estimate the core lateral spreading until the end of the plateau phase. We show it is expected to be limited and can thus be neglected, in both the ``de-beamed core plateau'' model presented in Sec.~\ref{sec:coreoffaxis}, and in the ``late deceleration plateau'' model described in Sec.~\ref{sec:losplateau}. However, only a detailed hydrodynamical simulation could fully confirm this assumption.

Sec.~ \ref{sec:losplateau}
We finally add an additional component to the light-curves at early times to represent the early steep decay (ESD) phase. This component is very commonly observed in early phases of GRB afterglows. It is often interpreted as originating from high-latitude emission of the material producing the prompt phase \citep{KP2000}, and is unlikely to be related to the forward shock. Due to relativistic beaming, it is typically dominated by material that is within several $\Gamma_0(\theta_{\rm v})^{-1}$ from the line of sight. We empirically model it here in the following way:
\begin{equation}
\label{eq:ESD}
    L_{\rm ESD}= \frac{E_{\gamma, \rm em}(\theta_{\rm v})}{T_{\rm 90}} \bigg(\frac{t}{T_{\rm 90}}\bigg)^{-3} \, ,
\end{equation}
where $T_{\rm 90}$ is the duration of the prompt emission phase and here, and in what follows we use a typical value of $T_{\rm 90}=20$\,s. This is a good approximation so long as the $\gamma$-ray energy is dominated by the line-of-sight material.

\section{Plateaus seen by off-core observers}
\label{sec:Plateauoffcore}
Relativistic beaming implies that material outside of an angle $\chi \approx 1/\Gamma$ from the line of sight contributes very little to the observed radiation. For observers at $\theta_{\rm v}>\theta_j$, where the energy content of the jet is smaller and the deceleration time longer, this implies a weaker prompt signal and an initially weak afterglow. As time goes by, the jet slows down and the observer can either (i) start receiving radiation from the more energetic material along the jet's core or (ii) simply start receiving significant contributions from material moving close to the line of sight that has eventually decelerated. 
Under certain conditions that we explore below, this can lead to a plateau-like phase in the X-ray light-curve. 
We turn next to a more in-depth description of these two possible plateaus that can be seen by observers outside the jet's core.

\subsection{Plateaus from jet core viewed off-axis}
\label{sec:coreoffaxis}
In the ``de-beamed core plateau" scenario presented here, the light-curve is dominated by material close to the core of the jet. In order to clearly separate this regime from the following one discussed in \S~\ref{sec:losplateau}, we focus on the case where the edge of the core is separated by an angle larger than $\Gamma_j^{-1}$ from the line of sight. In this regime, the most energetic part of the jet is initially beamed away from the observer due to relativistic beaming. The beaming decreases over time, until eventually the entire jet becomes visible. This results in a shallow plateau-like phase, assuming that $\Gamma_j^{-1}\ll \Delta \theta \lesssim 0.5\theta_j$ and a sufficiently large $\beta\gtrsim 8$ to ensure that the off-axis contribution from the core dominates that from material moving along the line-of-sight; (see 
Sec. 4.2. for details).
The duration of the plateau in this case is dictated by the time it takes the core to become visible to the observer, i.e., when $\Gamma(\theta_j)\approx \Delta \theta^{-1}$ or
\begin{eqnarray}
\label{eq:tplateaucore}
    & t_{\rm p}\approx t_{\rm d,los}(\theta_j)[ \Delta \theta\Gamma_j]^{1+2\epsilon \over \epsilon}\\
    &=\left\{ \begin{array}{ll}1700 E_{j,53}^{1/3} n_0^{-1/3} (\Delta \theta/0.02)^{8/3}\mbox{ s} & \mbox{ISM} ,\\
970 E_{j,53}A_{*,-1}^{-1} (\Delta \theta/0.02)^{4}\mbox{ s} & \mbox{wind} ,
\end{array} \right. \nonumber
\end{eqnarray}
where $E_j$ is the isotropic-equivalent kinetic energy\footnote{We adopt here and in what follows, the notation $q_x\equiv q/10^x$ in cgs units.
 of the blast wave at $\theta_j$ \citep{Granot2018}, }
\begin{equation}
E_j = \frac{\int_0^{\theta_j}\epsilon(\theta)2\pi\sin{\theta}\mathrm{d}\theta}{1-\cos{\theta_j}}\, .
\end{equation}

 The strong dependence on $\Delta \theta$ makes it easy to explain a wide range of plateau durations with little change in the viewing angle. Indeed, the observed distribution of $t_{\rm p}$ spans about three orders of magnitude (see, e.g., Figs.~\ref{fig:approxcorrelate}, \ref{fig:approxcorrelatelos}) which, assuming all other parameters are fixed, requires values of $\Delta \theta$ to vary by a factor of at most 13 for ISM, and 6 for wind. This is very reasonable given that the lowest value of $\Delta \theta$ in this scenario is $\Gamma_j^{-1}\approx 0.003$ and the largest is roughly $\theta_j/2\approx 0.05$.
If, in addition, one allows for variation in the core energy and ambient density, the same span of plateau durations can be reproduced with an even smaller range of $\Delta \theta$.

\begin{figure}
\centering
\includegraphics[width=0.38\textwidth]{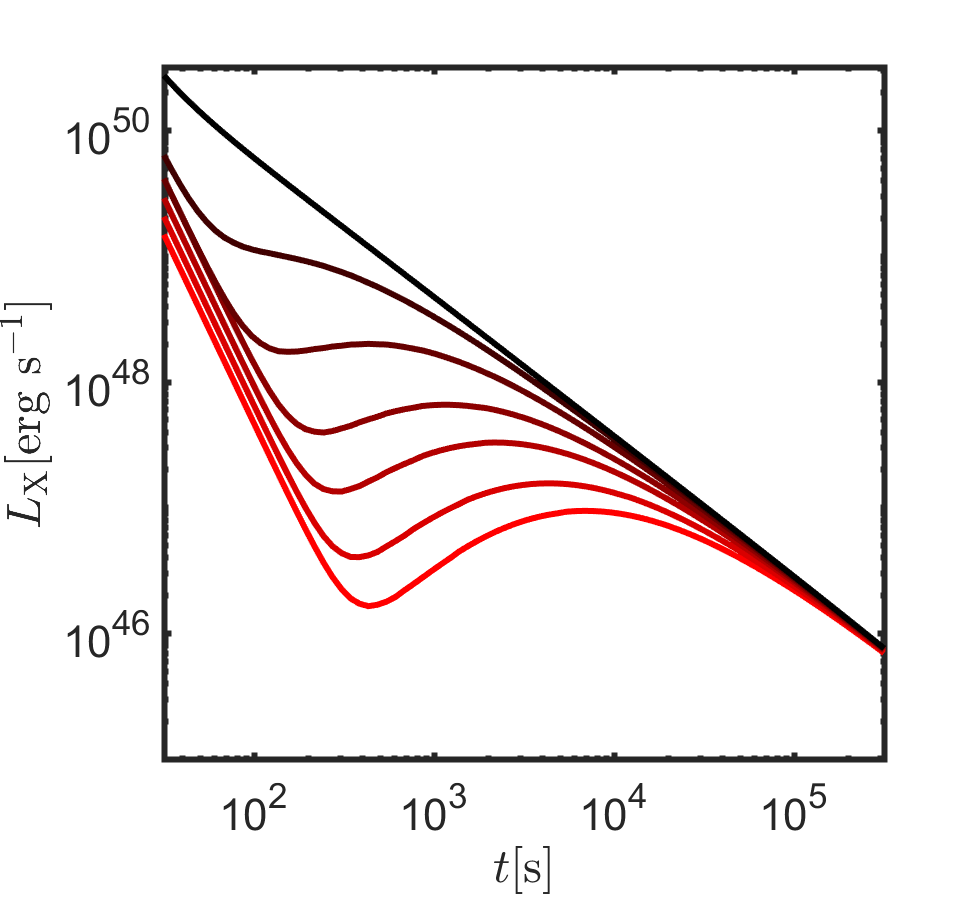}\\
\includegraphics[width=0.38\textwidth]{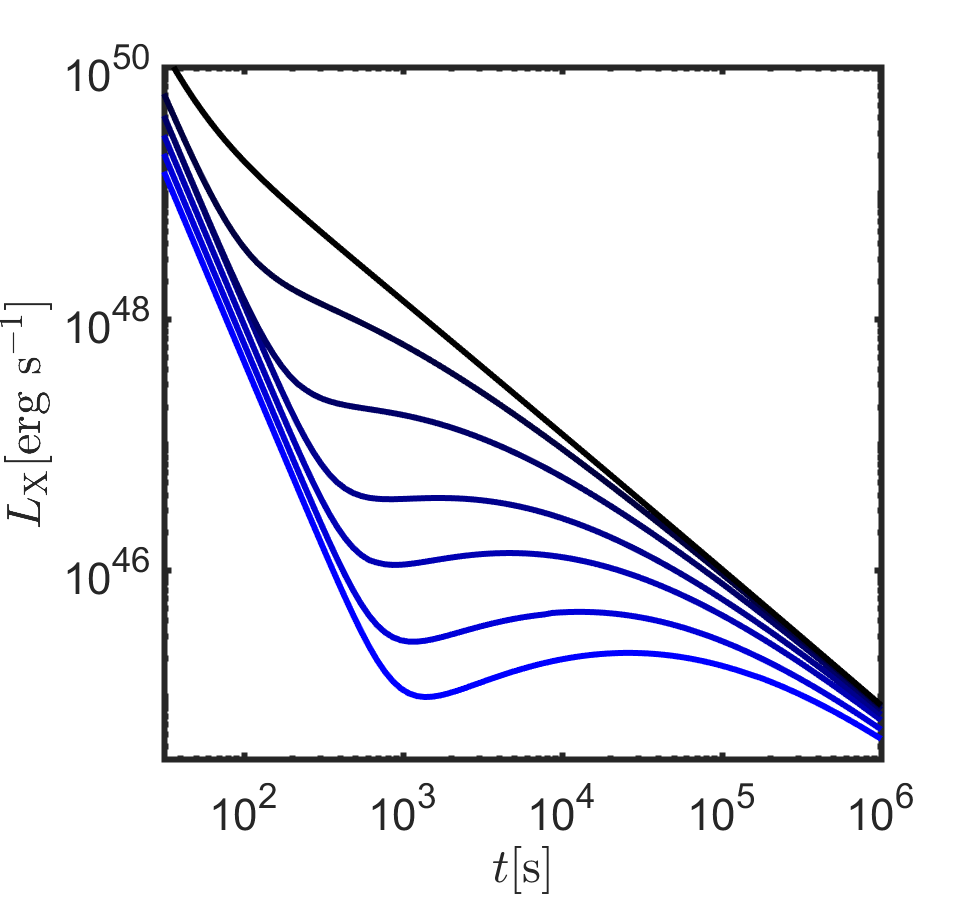}
\caption{De-beamed core plateaus: X-ray light-curves for a structured jet with $\alpha=8,\, \beta \gg 1$ (the latter is chosen to ensure that material from the core dominates the plateau, as described in \S~\ref{sec:coreoffaxis}) and different observation angles; from top to bottom: $\Delta \theta=0-0.03$ in steps of $0.005$. The X-rays are initially dominated by high-latitude emission, and at later times by the forward shock afterglow.
Results are shown for an ISM medium (top) with $n=1\,\mbox{ cm}^{-3}$ and a wind medium (bottom) with $A_*=0.1$. We have also taken here: $4\pi \epsilon_j= 10^{54}\,\mbox{erg}$, $\theta_j=0.1$, $\Gamma_j=400$,  $\eta_{\gamma}=0.1$, $\epsilon_e=0.1$, $\epsilon_B=0.01$, $p=2.2$.}
\label{fig:lightcurves}
\end{figure}

\begin{figure}
\centering
\includegraphics[width=0.38\textwidth]{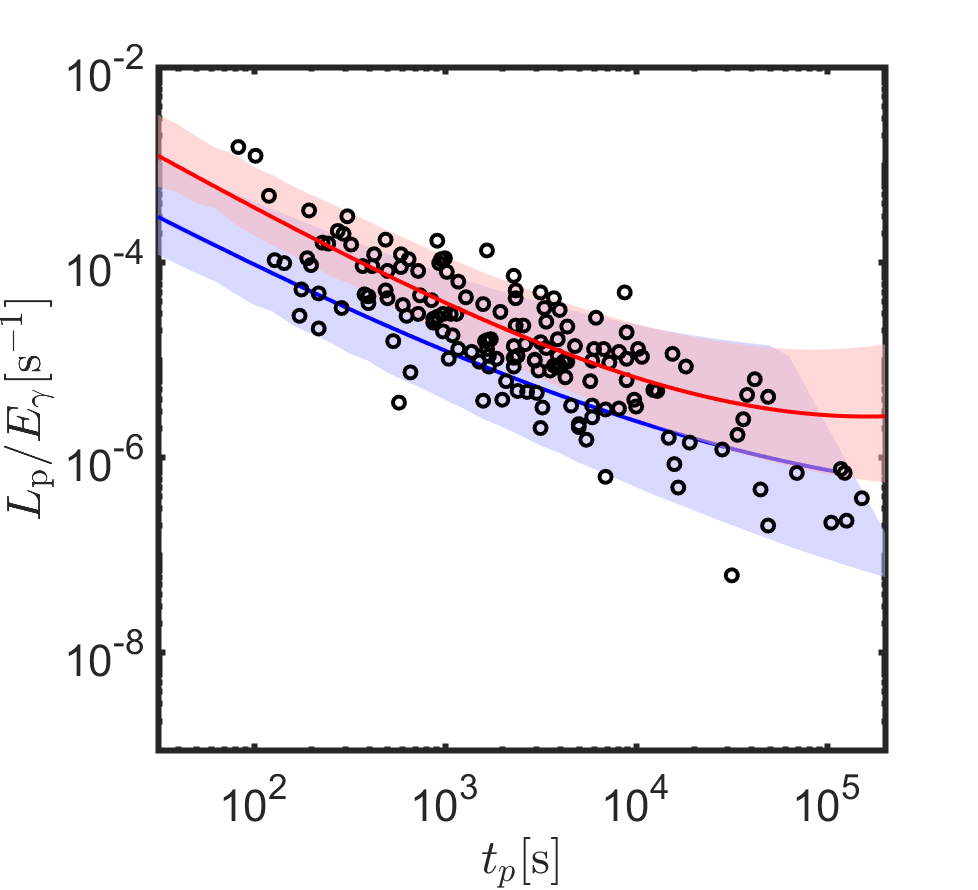}\\
\includegraphics[width=0.38\textwidth]{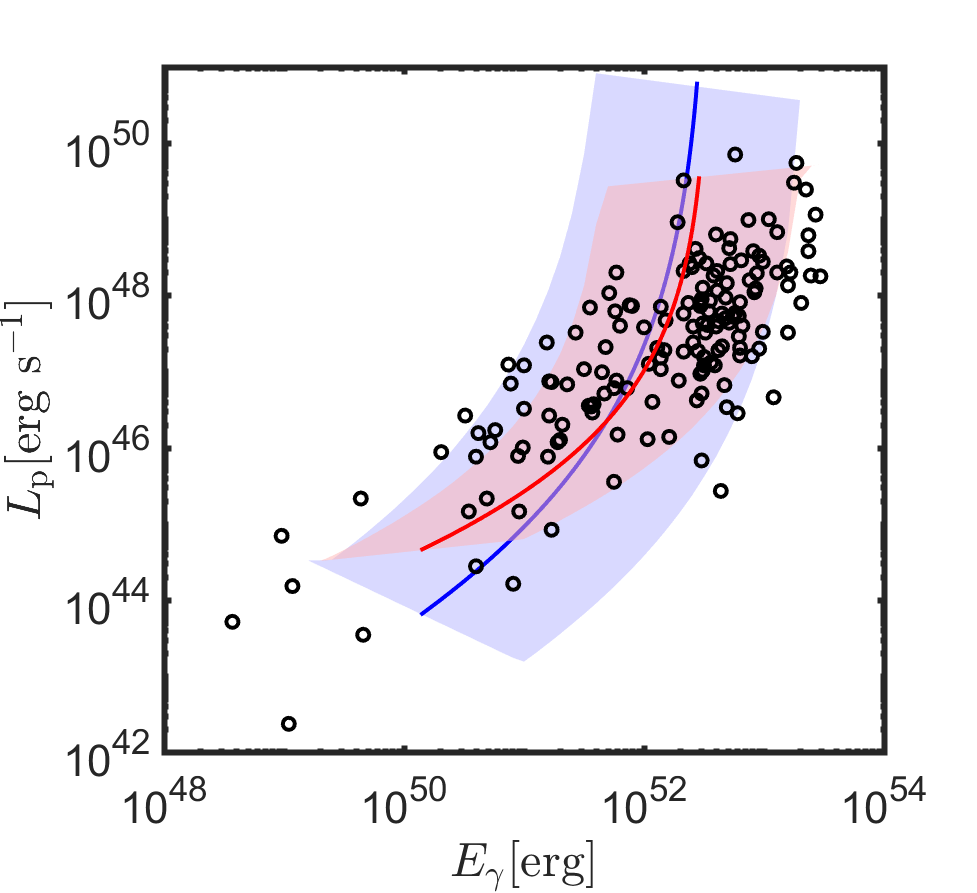}\\
\caption{De-beamed core plateaus: correlation between $L_{\rm p}/E_{\gamma}$ and $t_p$ (top)  and between $L_{\rm p}$ and $E_{\gamma}$ (bottom) as expected from Eqs.~\ref{eq:tplateaucore}, \ref{eq:Lplateaucore}, \ref{eq:threecore} for ISM (red) and wind environments (blue). Results are shown for $\alpha=8$, $\beta\gg 1$, $\theta_j=0.1$, $\eta_{\gamma}=0.05-0.2$, $p=2.2$, $4\pi \epsilon_j= 10^{53-54}\,\mbox{erg}$, $\Gamma_j = 400$, $\epsilon_e=0.1$, $\epsilon_B=0.01$ as well as $n=0.1-1\,\mbox{ cm}^{-3}$ for ISM ($A_*=0.1-1$ for wind). The solid lines depict the median choice of parameters in both cases, varying only the viewing angle and leaving all other parameters fixed. Circles mark observed GRB data, adapted from \citealt{Tang2019}.}
\label{fig:approxcorrelate}
\end{figure}

We turn next to calculate the luminosity at the end of the plateau phase.
The luminosity is somewhat reduced as compared to the standard on-axis case, in which the isotropic-equivalent energy of the jet's core is visible to the observer. This is because at $t=t_{\rm p}$, there is still a sizable fraction of the jet that lies beyond an angle of $\Gamma_j^{-1}$ from the observer, and its emission is therefore strongly suppressed. Since $\Gamma$ evolves slower in a wind environment, the effect is slightly more pronounced in that case. Naturally, regardless of the surrounding medium, at $t\gg t_{\rm p}$, when emission from the entire jet becomes visible, the luminosity seen by off-core observer matches that seen by on-axis observers.
The X-ray (defined here as $0.3-30$ keV) luminosity at the end of the plateau is therefore
\begin{eqnarray}
\label{eq:Lplateaucore}
& L_{\rm p}\!\approx\!  10^{46}f(1\!+\!z)^{2+p \over 4} \epsilon_{e,-1}^{p-1}\epsilon_{B,-2}^{p-2 \over 4} E_{j,53}^{2+p\over 4} t_{\rm p,3}^{2-3p \over 4}\bigg(\frac{4}{1+Y}\bigg)\mbox{ erg s}^{-1} \nonumber \\ 
&\approx\left\{ \begin{array}{ll} 7\times 10^{46}(1\!+\!z)^{2+p \over 4} \epsilon_{e,-1}^{p-1}\epsilon_{B,-2}^{p-2 \over 4} E_{j,53}^{2\over 3}\times \\ n_0^{3p-2\over 12}(\Delta \theta/0.02)^{4-6p\over 3}\frac{4}{1+Y}\mbox{ erg s}^{-1}& \mbox{ISM} ,\\ \\
1.5\times 10^{46}(1+z)^{2+p \over 4} \epsilon_{e,-1}^{p-1}\epsilon_{B,-2}^{p-2 \over 4} E_{j,53}^{2-p\over 2}\times \\ A_{*,-1}^{3p-2\over 4} (\Delta \theta/0.02)^{2-3p}\frac{4}{1+Y}\mbox{ erg s}^{-1} & \mbox{wind} ,
\end{array} \right.
\end{eqnarray}
where in the first line $f$ is a normalization that is 10 for ISM and 1.5 for wind. We have taken here the synchrotron spectrum branch above the cooling and injection frequencies $\nu_m$ and $\nu_c$, where the X-rays reside for typical jet parameters \citep{Nava2014,Santana2014,GvdH2014,Zhang2015,Beniamini2016,BvdH2017}. Also, $1+Y$ accounts for suppression due to SSC cooling and can be self-consistently calculated from  the other physical parameters and evolves very slowly with time (see \citealt{Beniamini2015} for details). For $p\approx 2.2$, one finds $(1+Y)\propto t^{-1/20}$ for ISM  and $(1+Y)\propto t^{-1/10}$ for wind. We have normalized it by the typical value obtained for the canonical parameter values chosen here (i.e., $\epsilon_e=0.1,\epsilon_B=0.01$ etc.).
For these values of $p$, one then finds that approximately $L_{\rm p}\propto t_{\rm p}^{-1}$ (see \S~\ref{sec:threeparam} for more details).
Light-curves arising from this scenario for a given set of physical parameters and changing values of $\Delta \theta$ are shown in Fig.~\ref{fig:lightcurves}.

Finally, the energy at the core of the jet can be related to the observed $\gamma$-ray energy in the prompt phase. Assuming that the observed $\gamma$-rays are always dominated by material moving along the line of sight (see \S~\ref{sec:modeling} for a justification of this), we have
\begin{equation}
\label{eq:Egammacore}
    E_{\gamma}=\frac{\eta_{\gamma}}{1-\eta_{\gamma}}E_{\rm k}(\theta_{\rm v})=\frac{\eta_{\gamma}}{1-\eta_{\gamma}}E_j \bigg(\frac{\theta_{\rm v}}{\theta_j}\bigg)^{-\alpha}.
\end{equation}
Writing $\theta_{\rm v}=\theta_j+\Delta \theta$ and using the relation between $\Delta \theta$ and $t_{\rm p}$ (Eq.~\ref{eq:tplateaucore}) we plug the previous expression into Eq.~\ref{eq:Lplateaucore} to obtain

\begin{align}
\label{eq:threecore}
& L_{\rm p}\!\approx \!10^{46}(1\!+\!z)^{2+p \over 4} \epsilon_{e,-1}^{p-1}\epsilon_{B,-2}^{p-2 \over 4} \bigg(\frac{1\!-\!\eta_{\gamma}}{\eta_{\gamma}} E_{\gamma, 53} \bigg)^{2+p\over 4}t_{\rm p,3}^{2-3p \over 4}\bigg(\frac{4}{1\!+\!Y}\bigg) \frac{\mbox{erg}}{\mbox{s}} \nonumber\\ & \times\left\{\! \!  \begin{array}{ll}7\left[1\!+\!0.16t_{\rm p,3}^{3/8}\bigg(\frac{1\!-\!\eta_{\gamma}}{\eta_{\gamma}}E_{\gamma,53}\bigg)^{-1/8}\! n_0^{1/8}\theta_{j,-1}^{-1}\right]^{(2+p)\alpha\over 4}\!&\! \mbox{ISM},\\ \\
1.5 \left[1\!+\!0.2t_{\rm p,3}^{1/4}\bigg(\frac{1\!-\!\eta_{\gamma}}{\eta_{\gamma}}E_{\gamma,53}\bigg)^{-1/4}\! A_{*,-1}^{1/4}\theta_{j,-1}^{-1}\right]^{(2+p)\alpha\over 4}\! &\! \mbox{wind}.
\end{array} \right.\!  %
\end{align}
The term in the bracket is the leading order approximation of $1+\left(\frac{t_{\rm p}}{t_j}\right)^{\epsilon \over 1+2\epsilon}$, where $t_j=t_{\rm d,los}(\theta_j)[ \theta_j\Gamma_j]^{1+2\epsilon \over \epsilon}$ is approximately the jet break time and $t_{\rm p}/t_j=(\Delta\theta/\theta_j)^{1+2\epsilon \over \epsilon}$. Writing the equation in this way makes it clear that since $\Delta \theta<\theta_j$, $t_p<t_j$. This means that the evolution immediately after the plateau still follows the normal pre-jet-break decline phase of GRB afterglows. For longer plateaus the two time-scales start approaching each other, leading to a shorter `normal decline' phase. In principle, a measurement of $t_p$, $t_j$ from observations of a given burst would lead to a direct estimate of $\Delta \theta/\theta_j$ that is independent of any of the other physical parameters.
However, as the viewing angle becomes larger, the jet break transition tends to become smoother, and so in practice it may prove quite challenging to extract this information from observations.

Eq.~\ref{eq:threecore} provides a relation between the three observable quantities $E_{\gamma}$, $L_{\rm p}$, $t_{\rm p}$ that is largely independent of the energy and Lorentz factor profile beyond the core. The correlation between  $L_{\rm p}/E_{\gamma}$ and $t_p$, as well as the correlation between $E_{\gamma}$ and $L_{\rm p}$ are depicted in Fig.~\ref{fig:approxcorrelate} as compared with observations. Note that the latter correlation does depend on the structure beyond the core.
It appears that the observed correlations can be readily reproduced. We stress that we do not attempt here any detailed fitting of the model, as there are clearly some degeneracies between some of the parameters which will hinder the usefulness of such an approach. The purpose of this figure is simply to demonstrate that correlations similar to the observed ones can naturally be reproduced by this model with very reasonable choices of the physical parameters.

We end this description by noting that this type of plateau will exist even in the idealized scenario of purely top hat jets, where there is no $\gamma$-ray and afterglow production by material beyond the core. In this case, the plateau properties remain the same as discussed above. However, in order for the $\gamma$-rays to remain detectable, the observation angle has to be somewhat closer to the core (i.e., $\Delta \theta \lesssim 5 \Gamma_j^{-1}$). $E_{\gamma}$ in Eq.~\ref{eq:Egammacore} is then obtained with the R.H.S. $(1 + q^{2})^{-2}$ term in Eq.~\ref{eq:Egamma}.

\subsection{Plateaus from material moving close to the line of sight}
\label{sec:losplateau}
In this case, the plateau is due to forward shock synchrotron emission from material travelling close to the line of sight that has not yet began decelerating significantly (``late deceleration plateau"). If the burst takes place in a wind environment, this scenario too can result in a plateau prior to the deceleration of the line-of-sight material (also, see \citealt{Shen2012}). The reason for this is that before deceleration, the energy in the forward shock scales as $E_{\rm k}\propto R \Gamma_0^2 \propto t$. Therefore, if the X-rays are above the cooling and injection frequencies (as expected, see \S~\ref{sec:coreoffaxis}), then $L_{\rm p}\propto E_{\rm k}^{(2+p)/4} t^{(2-3p)/4}\propto t^{(2-p)/2}$ which, for $p\approx 2.2$, is very close to being completely flat. This interpretation is in a way simpler than the previous one, as it could hold in principle even for bursts seen along their cores. However, it requires the deceleration peak, i.e., the break in the forward shock's component to the afterglow light-curve due to the onset of the deceleration of the emitting material, to occur at late times. As shown in Eq.~\ref{eq:tplatlos} below, unless the Lorentz factor of the emitting material is much smaller than expected for the jet core, this would not be easily achieved, without invoking very small values of the wind parameter, compared to theoretical expectations.

The duration of the plateau in this scenario is given by\footnote{Eq.~\ref{eq:tplatlos} holds for the `thin shell' approximation \citep{SariPiran1995}. The latter is valid so long as $t_{\rm d,los}\gg T_{90}$. Since by construction we are considering the situation in which $t_{\rm d, los}$ is long enough to power the plateau phase (lasting between hundreds to tens of thousands of seconds), this condition is expected to be satisfied.}
\begin{equation}
\label{eq:tplatlos}
    t_{\rm p}=t^{\theta_{\rm v}}_{\rm dec}(\theta_{\rm v}, 0)\approx 600 E_{j,53}A_{*,-2}^{-1}\Gamma_{j,2}^{-4}\bigg(\max\bigg[1,\frac{\theta_{\rm v}}{\theta_j}\bigg]\bigg)^{4\beta-\alpha}\mbox{ s}.
\end{equation}
Note in particular the small values of $\Gamma_j$ and of the wind parameter that were used above. Even with this choice, the plateau is barely noticeable beyond the ESD phase for GRBs seen within their central jet. 
Larger values would lead to shorter deceleration times and make the prospect of detecting this phase poorer still. If, however, $\theta_{\rm v}>\theta_j$, then (depending on $\alpha, \beta$) it could be possible to obtain plateaus, even with somewhat larger values of $\Gamma_j$ and/or $A_*$. For example, $\theta_{\rm v}=2\theta_j$ and $\alpha=\beta=3$ would stretch the duration of the plateau by a factor of $512$, which can be sufficient to lead to values of $t_{\rm p}$ that are comparable to observations. As in the previous interpretation, it is easy to explain a very wide range of plateau durations (as observed) by introducing rather small changes to $\theta_{\rm v}/\theta_j$.
An important difference between this and the de-beamed core scenario, is that here, as opposed to Eq.~\ref{eq:tplateaucore}, the Lorentz factor appears in the expression for the plateau duration. Furthermore, a combination of small $\Gamma_j$ and large $E_{j}$ is needed to have long plateaus in this scenario, which may be challenging to obtain in practice.

\begin{figure}
\centering
\includegraphics[width=0.38\textwidth]{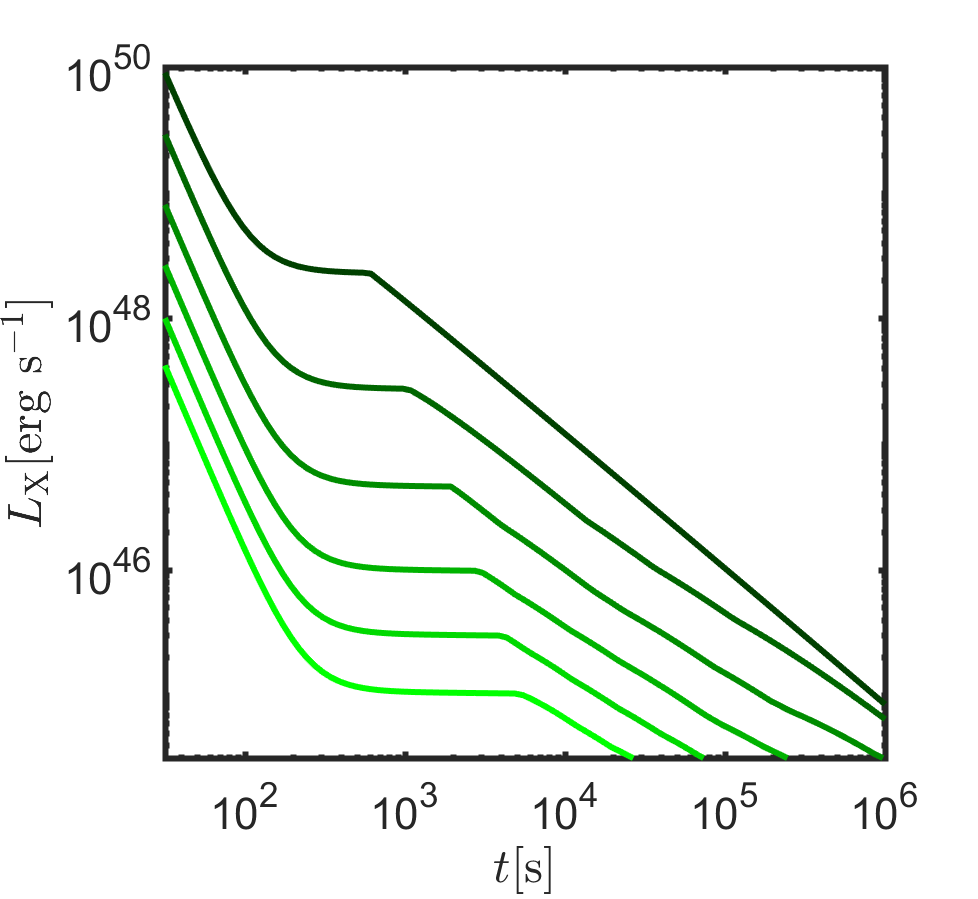}
\caption{Late deceleration plateaus: X-ray light-curves for a structured jet with $\alpha=8$, $\beta=3$ and different observation angles (from top to bottom: $\Delta \theta=0-0.1$ in steps of $0.02$). The X-rays are initially dominated by high-latitude emission, and at later times by the forward shock afterglow. The plateau in this case is produced by material moving close to the line of sight (and exists for a wind medium only). Results are shown for $4\pi \epsilon_j= 10^{54}\, \mbox{erg}$, $\theta_j=0.1$, $\Gamma_j=100$, $A_*=0.1$, $\eta_{\gamma}=0.1$, $\epsilon_e=0.1$, $\epsilon_B=0.01$, $p=2.2$. }
\label{fig:lightcurveslos}
\end{figure}

The luminosity at the end of the plateau is obtained by simply plugging $E_{\rm k}(\theta_{\rm v})$ into the standard forward shock synchrotron expressions. Assuming $\theta_{\rm v}>\theta_j$, we find
\begin{eqnarray}
\label{eq:Lplatlos}
  &  L_{\rm p}\!= 10^{47}(1\!+\!z)^{2+p \over 4} \epsilon_{e,-1}^{p-1}\epsilon_{B,-2}^{p-2 \over 4} E_{j,53}^{2+p\over 4} t_{\rm p,3}^{2-3p \over 4}\bigg(\frac{\theta_{\rm v}}{\theta_j}\bigg)^{-\alpha(2+p)\over 4} \nonumber \\ & \times\, \bigg(\frac{4}{1+Y}\bigg)\ \mbox{erg s}^{-1}\! \\ & \approx 10^{47} (1\!+\!z)^{2+p \over 4} \epsilon_{e,-1}^{p-1}\epsilon_{B,-2}^{p-2 \over 4} E_{j,53}^{4-2p\over 4}  \nonumber \\ & \times\, A_{*,-2}^{3p-2 \over 4}\Gamma_{j,2}^{3p-2} \bigg(\frac{\theta_{\rm v}}{\theta_j}\bigg)^{2\beta-3\beta p-\alpha+\frac{1}{2}\alpha p}\bigg(\frac{4}{1+Y}\bigg)\ \mbox{erg s}^{-1} .
\end{eqnarray}
As in \S~\ref{sec:coreoffaxis}, the approximate relation $L_{\rm p}\propto t_{\rm p}^{-1}$ is expected (see \S~\ref{sec:threeparam} for details). The plateau light-curves arising from this scenario are shown in Fig.~\ref{fig:lightcurveslos}.

The observed isotropic-equivalent prompt $\gamma$-ray energy is typically given by
Eq.~\ref{eq:Egammacore}, except for large viewing angles $\theta_\mathrm{v}\gg \theta_j$, or if the structure is very steep. Then, one can again easily relate the three observables $E_{\gamma}$, $L_{\rm p}$, $t_{\rm p}$,
\begin{eqnarray}
\label{eq:threelos}
 &   L_{p}\!=\!1.5\!\times\!10^{47}(1\!+\!z)^{2+p \over 4} \epsilon_{e,-1}^{p-1}\epsilon_{B,-2}^{p-2 \over 4} \bigg(\frac{1\!-\!\eta_{\gamma}}{\eta_{\gamma}} E_{\gamma, 53} \bigg)^{2+p\over 4} \nonumber \\ &  \times \, t_{\rm p,3}^{2-3p \over 4}\bigg(\frac{4}{1+Y}\bigg)\mbox{ erg s}^{-1}.
\end{eqnarray}
Similarly to the case in \S~\ref{sec:coreoffaxis}, the relationship given by Eq.~\ref{eq:threelos} is independent of the energy and Lorentz factor profiles beyond the core.

An independent correlation that can be compared with observations is the one between $L_{\rm p}$ and $E_{\gamma}$. For fixed burst parameters (and so long as $\eta_{\gamma}$ is independent of $\theta$) with varying viewing angles, we use Eqs.~\ref{eq:tplatlos}, \ref{eq:Lplatlos},  \ref{eq:Egammacore} to obtain $L_{\rm p}\propto E_{\gamma}^{1-\frac{p}{2}+\frac{\beta(3p-2)}{\alpha}}$. Since $1-\frac{p}{2}\approx 0$, and since the observed relation can be approximately fit by $L_{\rm p}\propto E_{\gamma}^X$, with an exponent $1\lesssim X \lesssim 1.5$, it is evident that, 
if the dominant effect explaining the observed $L_{\rm p}-E_{\gamma}$ correlation is a varying viewing angle from a burst to another, then $\alpha \gtrsim \frac{2}{3}(3p-2)\beta \approx 3\beta$ is needed in this model.
At the same time, Eq.~\ref{eq:tplatlos} clearly demands that $\alpha<4\beta$ in order for plateau durations to become more extended at larger viewing angles, which, in turn, is needed to obtain the values of some of the longer observed plateaus with realistic parameters. Therefore, barring possible inter-correlations between other burst parameters, some fine-tuning in this model is required to reproduce the observed $L_{\rm p}-E_{\gamma}$ correlation from viewing angle effects alone. Generally, a very steep structure is required for the profile of energy beyond the core.

\begin{figure}
\centering
\includegraphics[width=0.38\textwidth]{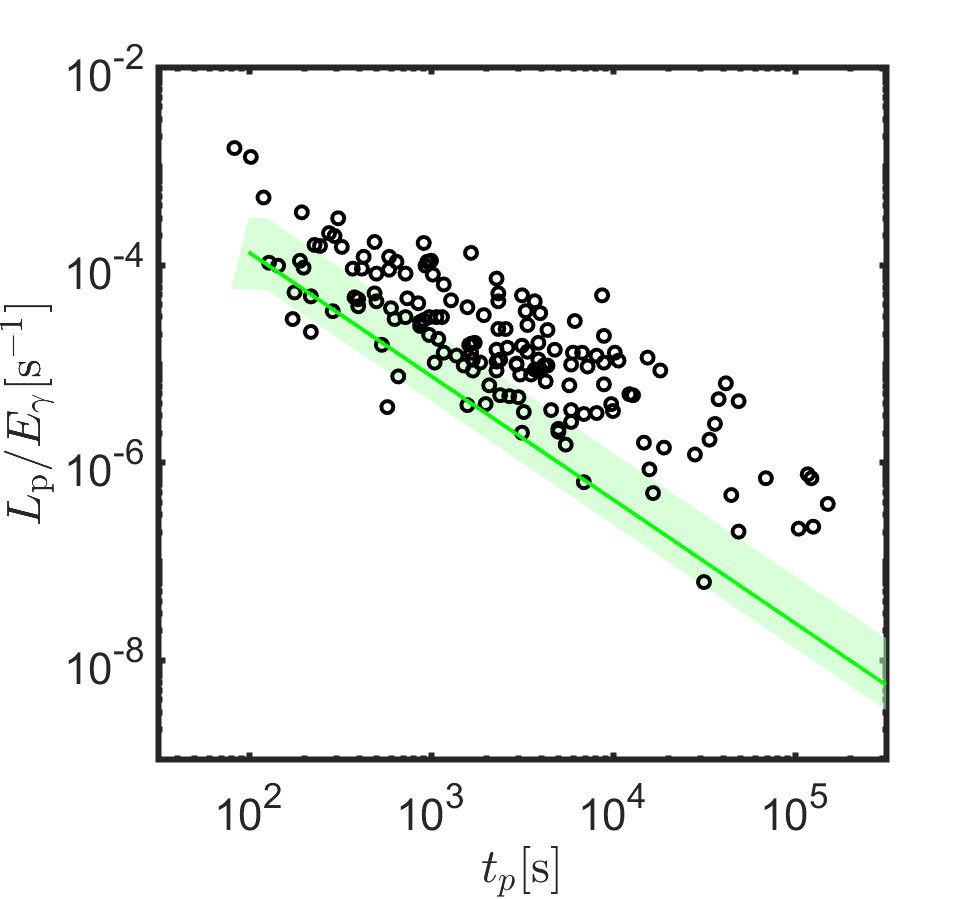}\\
\includegraphics[width=0.38\textwidth]{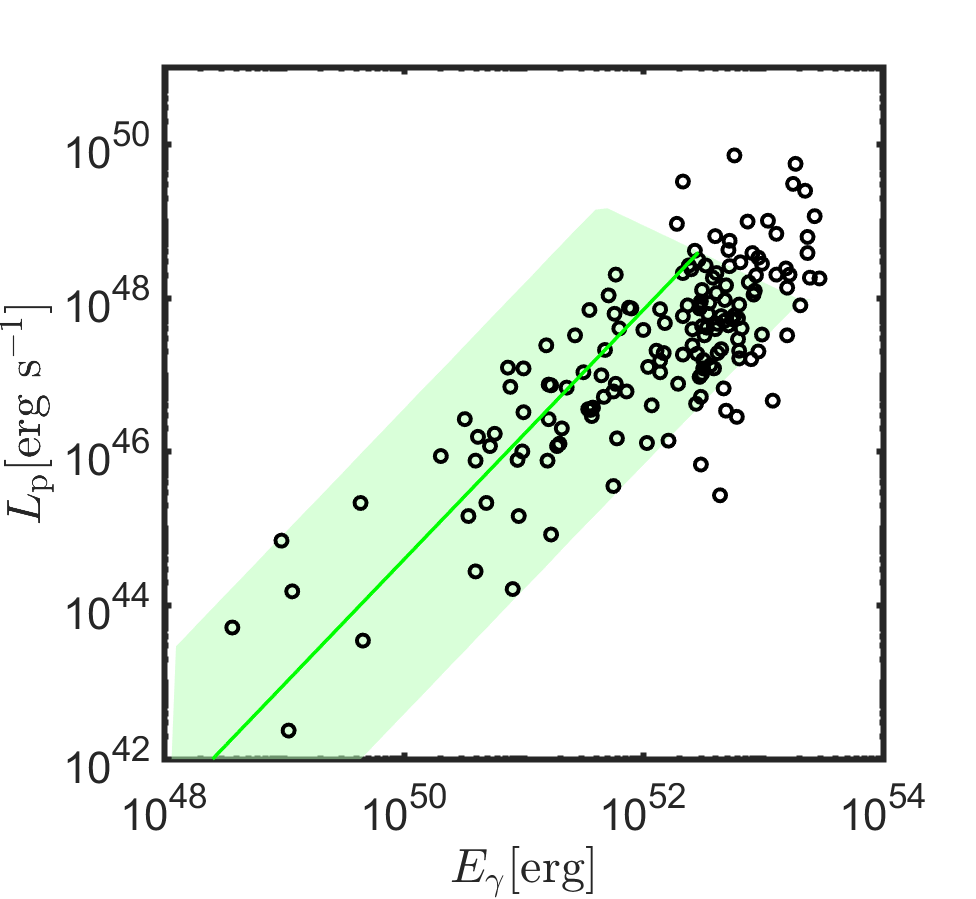}
\caption{Late deceleration plateaus: correlation between $L_{\rm p}/E_{\gamma}$ and $t_p$ (top)  and between $L_{\rm p}$ and $E_{\gamma}$ (bottom) as expected from Eqs.~\ref{eq:tplatlos}, \ref{eq:Lplatlos}, \ref{eq:threelos}. Results are shown for $\alpha=8$, $\beta=3$, $\theta_j=0.1$, $\eta_{\gamma}=0.05-0.2$, $\Gamma_j=100$, $4\pi \epsilon_j= 10^{53-54}\,\mbox{erg}$, $\epsilon_e=0.1$, $\epsilon_B=0.01$, $p=2.2$, $A_*=0.1-1$ for wind. The solid lines depict the median choice of parameters, varying only the viewing angle and leaving all other parameters fixed. Circles mark observed GRB data, adapted from \citealt{Tang2019}. }
\label{fig:approxcorrelatelos}
\end{figure}

The model as well as the observed correlations are depicted in Fig.~\ref{fig:approxcorrelatelos}.
The $L_{\rm p}/E_{\gamma}- t_p$ relation may be approximately reproduced, while, as mentioned above, the $L_{\rm p} - E_{\gamma}$ correlation requires a rather steep energy structure beyond the core as compared with the Lorentz factor structure, which is in some tension with the requirement for producing long-lived plateaus in this scenario.

\section{Discussion}
\label{sec:Discuss}
\subsection{Relationships between plateau and prompt properties}
\label{sec:threeparam}
We begin this section by noting on a commonality between the two types of plateaus explored in \S~\ref{sec:coreoffaxis}, \ref{sec:losplateau}, which will indeed persist in any interpretation within which there is a strong correlation between the prompt $\gamma$-ray energy and the kinetic energy used to power the plateau which is largely independent of the energy and Lorentz factor profiles beyond the jet core. This commonality has to do with a specific relationship between the three observable parameters: the isotropic-equivalent $\gamma$-ray energy ($E_{\gamma}$) the duration of the plateau ($t_{\rm p}$) and the luminosity at the end of the plateau ($L_{\rm p}$). Let us assume that $E_{\gamma}\propto E_{\rm k}$ where $E_{\rm k}$ is the kinetic energy used to power the plateau.
Under the usual Blandford-Mckee blast wave evolution,  $E_{\rm k}$ is tapped to radiation mainly through the forward shock as the blast wave interacts with the surrounding medium.
As stated in \S~\ref{sec:coreoffaxis}, the X-rays reside above $\nu_c$, $\nu_m$ for typical burst parameters. Thus, the luminosity scales with the kinetic energy and the time as: $L_{\rm p}\propto E_{\rm k}^{(2+p)/4}t_{\rm p}^{(2-3p)/4}\propto E_{\gamma}^{(2+p)/4} t_{\rm p}^{(2-3p)/4}$. For $p\approx 2$, this leads to $L_{\rm p}\propto E_{\gamma}t_{\rm p}^{-1}$ which is close to the observed relation.
Some small modifications to the relation above are expected due to the effects of, e.g., {\it slight} deviations from the linear relation between $E_{\gamma}$ and $E_{\rm k}$ assumed above (as in \S~\ref{sec:coreoffaxis}) and SSC cooling effects, causing a slightly shallower evolution of the luminosity with time  (see \citealt{Beniamini2015}).
Note, however, that this correlation is much less natural in the common interpretation of plateaus that associates them with large amounts of energy injection onto the external shock at late times. In the latter interpretation, the available energy at the time of $\gamma$-ray production is much smaller \footnote{This also often leads to uncomfortably large values of the prompt gamma-ray efficiency.} than, and not necessarily correlated with, the kinetic energy of the blast-wave at the end of the energy injection phase, and the reasoning above will no longer hold.

\subsection{Differentiating between plateau origins}
\label{sec:differences}
It is plausible that both plateau origins discussed in this paper manifest in some cases. Indeed the fact that both possibilities can adequately explain the observed correlations between $E_{\gamma}$, $L_{\rm p}$, $t_{\rm p}$, could make distinguishing between them a challenging endeavour. Nonetheless, it is interesting to test whether there exist some (possibly more detailed) observational tests to compare these (and other) models.

One difference between the two models that is clear from Figs.~\ref{fig:lightcurves}, \ref{fig:approxcorrelatelos} is the shape of the light-curves in both cases during the plateau. The de-beamed core model can result in a range of behaviours, from slowly declining plateau phases to ones that exhibit a shallow bump. Indeed this kind of behaviour is observed in some cases (of the order of a few percent of the entire population). Some examples
within \textit{Swift} bursts are: GRBs 081028, 090205, 100901A, 110213A, 120118B, 120215A, 120224A, 150911A, 170202A, 170822A, 181110A, 190422A. The late deceleration plateaus, on the other hand, exhibit a roughly universal plateau phase, that is almost completely flat. Although this is consistent with some GRB observations, this does not seem to apply to all or even most observed plateaus.

Another major difference in the physical set-up leading to the two types of plateaus discussed here regards the angular profile of the Lorentz factor of the jet beyond the jet core. The de-beamed core model requires relatively small values $\Delta \theta$ as well as a profile of $\Gamma$ that falls rapidly beyond the core to avoid the afterglow from the line-of-sight material from dominating over the off-axis contribution from the core. Typically, $\beta \gtrsim 8$ is required. However, this steepness does not necessarily turn off the emission from line-of-sight material along viewing angles which are considered here. For example for $\Gamma_{j}=400, \beta = 8, \Delta \theta =0.03$, this material still has a sizeable initial Lorentz factor of $\Gamma_0(\theta_v)\gtrsim 50$ and contributes to the afterglow. 

Alternatively, the other type of plateaus discussed here, due to late deceleration of line-of-sight material require smaller values of $\Gamma_j$ and somewhat shallower angular profiles of the Lorentz factor. Furthermore, as shown in \S~\ref{sec:losplateau}, in order to reproduce the observed $L_{\rm p}-E_{\gamma}$, a rather steep energy structure beyond the core (with $\alpha\approx 3\beta$) is needed.

In the new era of gravitational waves (GW) detections of short GRBs, we now have the possibility to observe (and measure) large viewing angles of GRBs. We thus may be able to resolve these different possibilities for the jet structure and prompt emission at large latitudes, by collecting statistical data on the properties of the prompt \citep{Beniamini2019} and afterglow \citep{Gottlieb2019,Duque2019} emission of such bursts. This in turn could potentially distinguish between the plateau scenarios discussed here. It should be noted, however, that plateaus are more often observed in long GRBs (see \citealt{Margutti2013} and \S~\ref{sec:shortGRBs} below), and it remains an open question whether or not the structure of short and long GRB jets are similar.

In the case of GRB170817A, currently the event with the most detailed insight into the structure of any GRB jet, afterglow photometry points towards a somewhat more shallow jet structure, with $\alpha, \beta \gtrsim 2$ \citep[e.g.,][]{Gill2018,Ghirlanda2019}. This, however, does not contradict the aforementioned requirement of our de-beamed core plateau model for rather large $\beta$-values. Indeed, on the one hand, the ability to produce core de-beamed plateaus is sensitive to the jet's structure only up to near-core angles (recall that, here, we consider $\Delta \theta < \theta_j \approx 0.1~{\rm rad}$). On the other hand, the afterglow photometry of far-core events, such as GRB170817A, is sensitive to the overall structure, up to $\theta \gtrsim 25~{\rm deg}$. In other words, a jet structure with a sharp drop ($\alpha, \beta \gtrsim 8$) between $\theta_j$ and $1.5\theta_j$, and then a shallow decrease ($\alpha, \beta \gtrsim 2$) for larger angles is compatible with both a plateau phase for near-core observers and a GRB170817A-type afterglow for far-core observers.

We end this discussion with two slightly more speculative directions of investigation that may help distinguish between the plateau models.
The first involves the reverse shock emission. Since the reverse shock's behaviour before and after the deceleration time is very distinct--as seen from material traveling along the line of sight to the observer--, one may be able to test the late deceleration plateau scenario, by searching for GRBs in which there are both an observed reverse shock emission and a plateau phase. The reverse shock component, if present, should evolve significantly from before to after $t_{\rm p}$. However, this approach may be hindered in practice since the reverse shock contribution can rarely be identified with confidence and indeed may be extremely weak if the GRB ejecta is even moderately magnetized.

The second avenue of exploration regards the polarization of the plateau.
A full analysis of the polarization signal from these models is rather involved and deserves a more detailed study elsewhere. Furthermore, it requires some additional assumptions, e.g., regarding the symmetry of the magnetic fields in the plane of the shock. Nonetheless, we mention here in passing that we may expect to have different polarization signatures in the two scenarios discussed here. In the de-beamed core model, there is a preferred orientation of the emitting material relative to the observer, which could result in a polarized signal, while in the late deceleration model the emitting material is roughly symmetric around the line of sight and the overall polarization signal would likely be much smaller.

\subsection{Plateau statistics}
It is illuminating to consider also that the fraction of bursts that exhibit plateaus is $\approx 0.5$ \citep{Kumar2015}. Under the de-beamed core model interpretation, this can easily be related to the maximal angle at which cosmological bursts can typically be viewed, $\theta_{\rm max}$. A reason that such a limiting observation angle exists can be due to a strong reduction in the $\gamma$-ray producing efficiency beyond the core of the jet (see \citealt{BN2019,Gottlieb2019b} for more details).

Unless $\epsilon$ has decreased significantly by $\theta_{\rm max}$, the fraction of bursts with plateaus is roughly proportional to the solid angle of on- and off-axis observable bursts, i.e.,
\begin{equation}
    \frac{\theta_{\rm max}^2-\theta_j^2}{\theta_{\rm max}^2}\approx 0.5 \, ,
\end{equation}
or $\theta_{\rm max}\approx 1.4\theta_j$.
In other words, since there is a significant fraction of bursts with {\it no} plateaus, the maximum angle at which cosmological bursts can be detected cannot be much larger than the jet opening angle, $\theta_j$. This is consistent with the results of \cite{BN2019} mentioned above. Furthermore, note that this argument becomes even more stringent if some of the plateaus are not due to the off-axis origin.

Since, in the late deceleration model, plateaus may appear even for on-axis observers, it is less straight-forward to use the plateau statistics to constrain the viewing angle in this case. Nonetheless, since the energy structure in this scenario must be very steep (see \S~\ref{sec:losplateau}), this limits how large the viewing angles of typical cosmological GRBs can be before they become undetectable.

\subsection{Spectrum and appearance at other wavelengths}
\label{sec:wavelengths}
Observationally, there is usually no evidence for a change in spectrum between the plateau phase and the following X-ray emission \citep{Kumar2015}. For scenarios in which the plateau is produced internally, i.e., where the emission is from material dissipating at radii smaller than the external shock (see \S~\ref{sec:Intro} for examples) this requires fine tuning and should therefore be a source of concern regarding their viability for producing {\it the majority} of the observed plateaus. In both of the scenarios proposed here, the cause for the end of the plateau is geometric or dynamical in nature, and therefore there is no change of spectrum associated with the plateau's demise.

Extending beyond the X-rays, it is interesting to consider optical observations simultaneous to X-ray plateaus. As it turns out, the observed situation is somewhat complex \citep{Panaitescu2006,Li2012,Liang2013}. In some cases, there are simultaneous plateaus in optical and X-rays, while in others the optical band exhibit a distinct temporal behaviour to the X-rays. 
In both of our models, the optical may either mimic the X-rays or not, depending on the location of the injection ($\nu_m$) and cooling ($\nu_c$) frequencies at the time of the plateau. With reasonable variations in the microphysical parameters, it is quite possible for the optical band to be, in some cases, in between $\nu_m$, $\nu_c$ during an X-ray plateau, while in others, to be above both frequencies.

In particular, for the late deceleration plateaus there are two extra possibilities on top of the case $\nu_{\rm obs}>\nu_m,\nu_c$ that was already explored above. First, $\nu_c<\nu_{\rm obs}<\nu_m$. Here $L\propto E^{3/4} t^{-1/4}$, so that as long as $E\propto t$ (see \S~\ref{sec:losplateau}), we get $L\propto t^{1/2}$. Second, we consider $\nu_m<\nu_{\rm obs}<\nu_c$. In this case, in a wind medium, $L\propto E^{1+p\over 4}t^{1-3p\over 4}$ leading to $L\propto t^{2-2p \over 4}\approx t^{-1/2}$. That is, the optical flux could be either rising or declining during the X-ray plateau. Since in a wind environment $\nu_c$ increases over time, and $\nu_m$ decreases, a typical progression is $\nu_c<\nu_{\rm obs}<\nu_m$ leading to $\nu_c,\nu_m<\nu_{\rm obs}$ and finally $\nu_m<\nu_{\rm obs}<\nu_c$. Which one or more of these intervals will be seen in the optical during an X-ray plateau depends on the microphysical parameters and the viewing angle and could therefore lead to quite a complex relationship between the simultaneous observations in both bands.

A somewhat analogous situation arises in the de-beamed core model, although there the energy increase over time is not due to more matter being decelerated, but rather due to more matter contributing towards the line of sight. In this case, the relationship between the two is no longer given by a simple power-law scaling. Note that, in the ISM case, $\nu_c$ decreases over time, so the ordering $\nu_m<\nu_{\rm obs}<\nu_c$ does not occur at late times like in the wind case.

\begin{figure}
\centering
\includegraphics[width=0.35\textwidth]{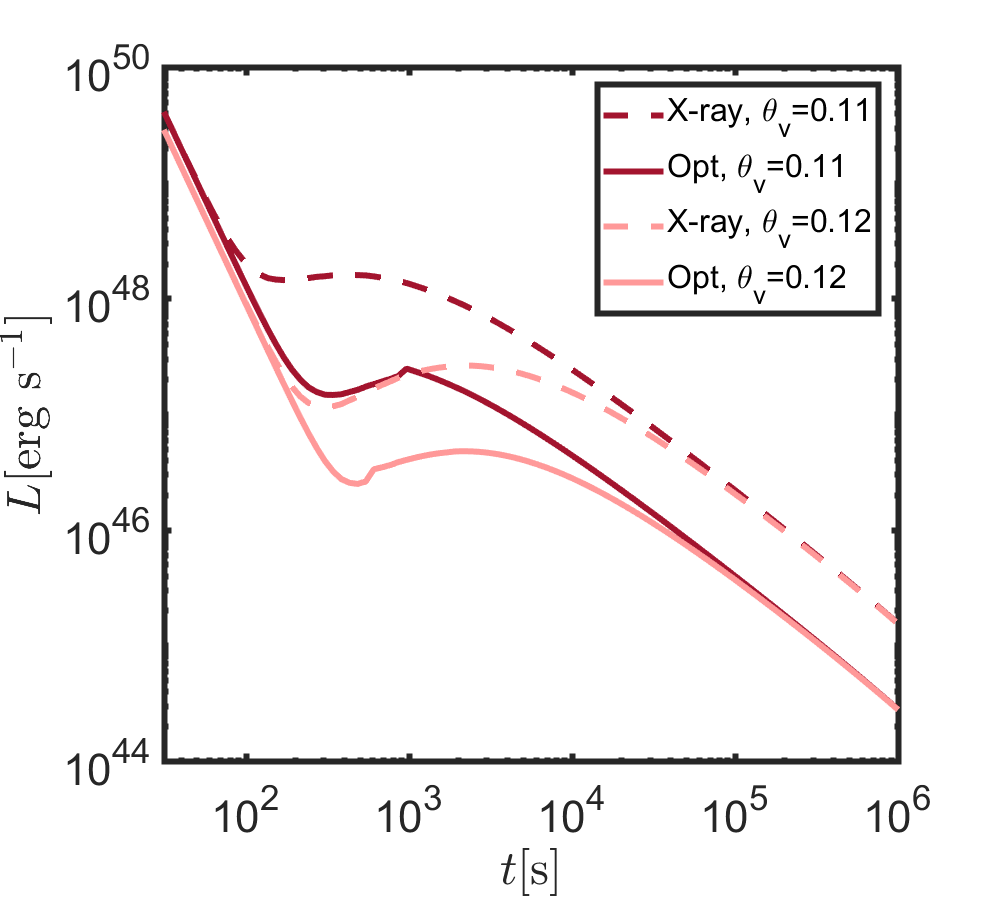}\\
\includegraphics[width=0.35\textwidth]{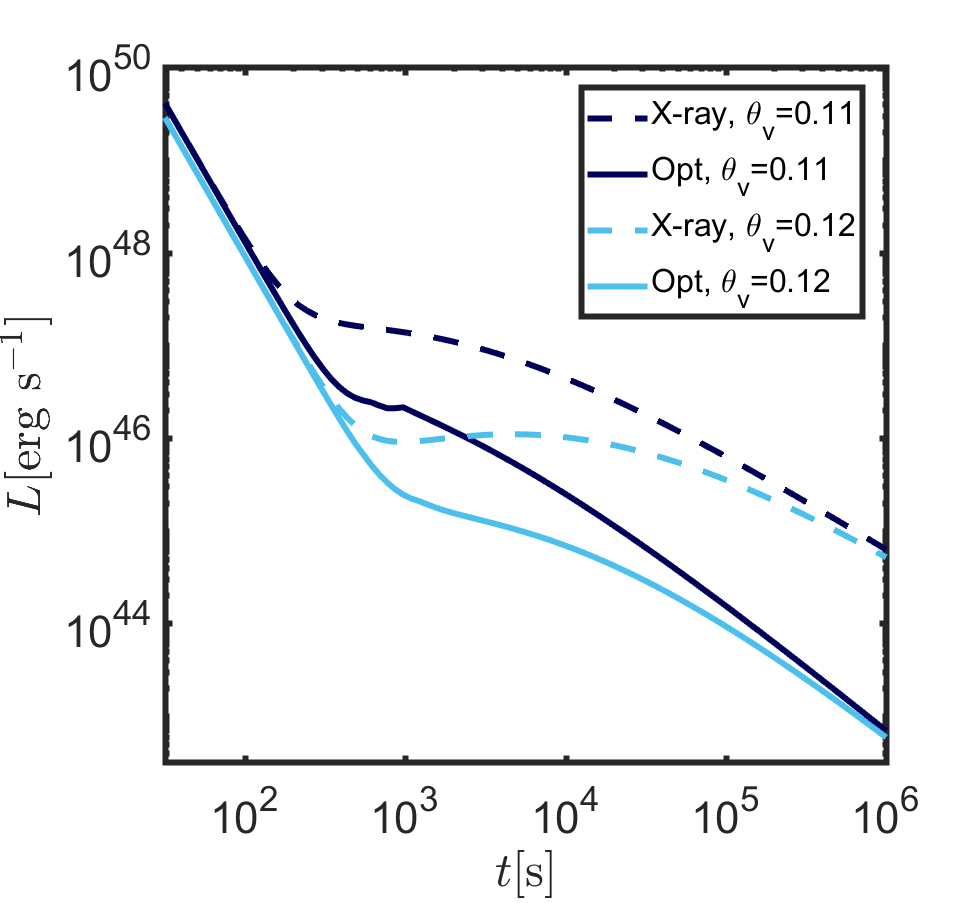}
\caption{De-beamed plateaus: X-ray (solid) and optical (dashed) light-curves for a structured jet with $\alpha=8$, $\beta\gg 1$, $\theta_j=0.1$, $\eta_{\gamma}=0.1$, $\Gamma_j=400$, $4\pi \epsilon_j= 10^{54}\,\mbox{erg}$, $\epsilon_e=0.1$, $\epsilon_B=0.01$, $p=2.2$ as well as $n=1\,\mbox{ cm}^{-3}$ for ISM (red) and  $A_*=0.1$ for wind (blue). For ease of comparison with observations we use here a 0.3-10 keV range for the X-rays and the R band for the optical.}
\label{fig:Xrayoptcore}
\end{figure}

\begin{figure}
\centering
\includegraphics[width=0.35\textwidth]{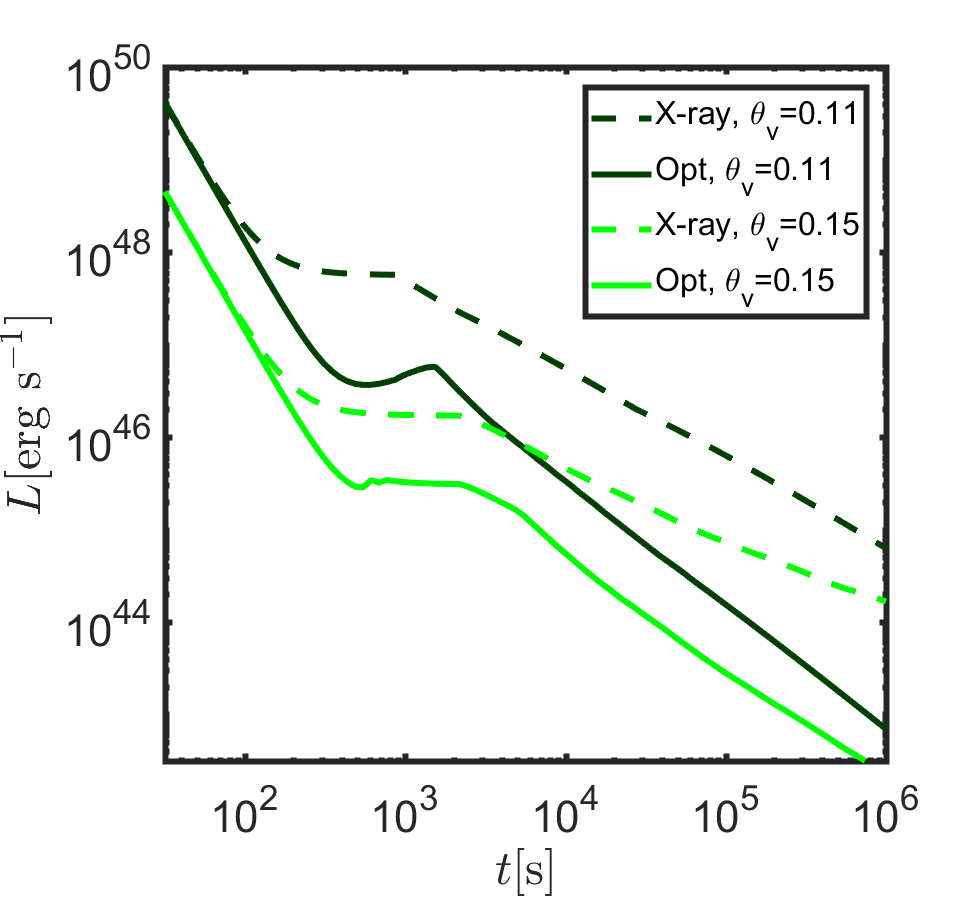}
\caption{Late deceleration plateaus: X-ray (solid) and optical (dashed) light-curves for a structured jet with $\alpha=8$, $\beta=3$, $\theta_j=0.1$, $\eta_{\gamma}=0.1$, $\Gamma_j=100$, $4\pi \epsilon_j= 10^{54}\,\mbox{erg}$, $\epsilon_e=0.1$, $\epsilon_B=0.01$, $p=2.2$, $A_*=0.1$. For ease of comparison with observations we use here a 0.3-10 keV range for the X-rays and the R band for the optical.}
\label{fig:Xrayoptlos}
\end{figure}

A side-by-side comparison of X-ray and optical light-curves that will be seen for given GRBs, with different physical parameters and viewing angles is shown in Figs.~\ref{fig:Xrayoptcore}, \ref{fig:Xrayoptlos}.
The optical light curves are computed here using the standard prescription from \citet{Sari1998}, i.e., assuming that all electrons in the shocked external medium have the same properties ($\gamma_\mathrm{m}$, $\gamma_\mathrm{c}$, etc.) as freshly accelerated electrons at the shock front. This approximation is less justified for slow cooling electrons, especially in the wind case where $N_\mathrm{e}\propto R$ \citep[see, e.g.,][]{Beloborodov2005}. Taking into account a more realistic treatment of the evolution of electrons in the shocked region would smooth the optical light-curves and enhance chromatic behaviours \citep[see, e.g.,][]{Uhm2012}.  

\subsection{Plateaus in short GRBs}
\label{sec:shortGRBs}
Although less frequent, plateaus are also observed in short GRBs. An examination of the {\it Swift} database\footnote{\url{https://swift.gsfc.nasa.gov/archive/grb_table/} \citep{Gehrels2004}} suggests that when a plateau is seen in a short burst it often has a short duration. Indeed, from the plateau duration-luminosity relation, short plateaus have a larger luminosity and therefore long plateaus might be observationally discriminated against because they are weaker. Furthermore, assuming typical va\-lues for the isotropic-equivalent kinetic energy and external density in long vs. short GRBs we find that, for the same duration, the plateau luminosity is weaker in short GRBs. To illustrate the latter point, consider the de-beamed core model. Taking $E_{\rm j}=10^{51}$\mbox{ erg} ($E_{\rm j}=10^{53}$\mbox{ erg}) and $n=0.1\mbox{ cm}^{-3}$ ($A_*=0.1$) as typical values for short and long bursts respectively, Eq.~\ref{eq:tplateaucore} results in comparable durations, 
\begin{eqnarray}
& t_{\rm p,S}=800 E_{S,51}^{1/3} n_{-1}^{-1/3} (\Delta \theta/0.02)^{8/3}\mbox{ s} , \nonumber \\
& t_{\rm p,L}=970 E_{L,53}A_{*,-1}^{-1} (\Delta \theta/0.02)^{4}\mbox{ s} ,
\end{eqnarray}
where the sub-script $S$ ($L$) denotes short (long) GRBs. Using Eq.~\ref{eq:Lplateaucore} we can obtain the ratio of the plateau luminosities for the same parameters
\begin{equation}
\label{eq:Lpratio}
\frac{L_{\rm p,S}}{L_{\rm p,L}}=0.08 E_{\rm S,51}^{2/3}E_{\rm L,53}^{1/10}n_{-1}^{0.38}A_{*,-1}^{-1.15}(\Delta \theta/0.02)^{1.53} \, ,
\end{equation}
where, for clarity, we have used here a typical value of $p=2.2$.
Eq.~\ref{eq:Lpratio} demonstrates that the ratio is small for typical values of the burst parameters, making plateaus of a given duration more faint in short as compared with long GRBs. Notice that if the external density in the vicinity of short GRB explosions is weaker (as may be expected for double neutron stars mergers with strong kicks or delays between formation and merger), the conclusion regarding the luminosity ratio becomes even stronger. Naturally, one should also take into account the difference in typical distances between short and long bursts. Since short GRBs are likely to on average be closer than long GRBs, the ratio of the observed fluxes might be somewhat closer to unity as compared to the luminosity ratio. 
Still, this is unlikely to qualitatively change the conclusion\footnote{As an illustration changing the typical redshift between $z=1$ for short bursts and $z=2$ for long bursts, corresponds to a modification by a factor of $\lesssim 6$ between the luminosity ratio in Eq.~\ref{eq:Lpratio} and the corresponding flux ratio.}. Overall, short GRB plateaus, and especially the longer ones, are expected to be harder to detect than those of long GRBs.

\section{conclusions}
\label{sec:conclusion}
We have presented here an interpretation of X-ray plateaus, linking them to forward shock emission viewed by observers on lines of sight very slightly beyond the GRB jet's core. Depending on the jet structure, such observers may see a plateau in the early X-ray afterglow light-curve that is either due to de-beamed emission from the core coming gradually into view or else from material travelling close to the line of sight that has not yet decelerated significantly.
The latter interpretation requires a wind-like medium and although it could in principle hold also for observers along the jet's core, it requires extreme choices of the physical parameters to be realized in those cases, and instead, is more easily seen by off-axis observers. Due to the strong dependence on viewing angle, both interpretations can reproduce the large span of observed plateau durations and luminosities with very modest variations in the viewing angle between bursts. Furthermore, they can reproduce the observed correlations between the isotropic-equivalent $\gamma$-ray energy ($E_{\gamma}$) the duration of the plateau ($t_{\rm p}$) and the luminosity at the end of the plateau ($L_{\rm p}$).
We note, however, that the late deceleration model requires more fine tuning of the energy and Lorentz factor structures in order to do so and also results in a roughly universal (almost completely flat) evolution of the X-ray light-curve before the end of the plateau, which is less commonly observed. As such, it appears more likely that this scenario manifests in some, but perhaps not the majority, of observed plateaus.

Generalizing beyond the two models studied in this work, we have shown that the observed correlations arise in models where $E_{\gamma}$ is roughly linearly correlated with, and represents a large fraction of, the blast wave kinetic energy tapped during the plateau phase. The most common interpretation for the plateau, involving significant injection onto the external shock at late times does not naturally reproduce these properties.

Due to the geometric and dynamical interpretations associated with these plateau models, no spectral change is expected between the plateau and post-plateau emission. In the optical band, due to interplay with the characteristic synchrotron frequencies, complex behaviours are possible. This feature is consistent with the observation that during the time of X-ray plateaus, the optical light-curves of the same GRBs are in some instances also flat, but in others are not.

The fraction of bursts that exhibit plateaus, and the statistics of their durations can be related in these models to the distribution of viewing angles. Indeed, the fact that only $\approx 0.5$ of bursts have an X-ray plateau is consistent with the interpretation that cosmological bursts are viewed at most only slightly off-core. The latter point is both required by multiple lines of evidence from observations \citep{BN2019} and is natural in various prompt emission models, that will lead to very inefficient $\gamma$-ray production at angles beyond the core where the energy or the Lorentz factor have significantly decreased.

Finally, determining which (if any) of the plateau interpretations presented here is dominant, could be aided by observational constraints on the energy and Lorentz factor structures beyond the jet's core. Indeed, these are expected to be probed, at least for short bursts, correlations in the near future with the advent of GW-triggered GRBs \citep{Beniamini2019,Gottlieb2019,Duque2019}. 

\section*{Acknowledgements}
We thank Jonathan Granot, Ehud Nakar, Pawan Kumar and Maria Dainotti for helpful discussions. P. Beniamini's research was funded in part by the Gordon and Betty Moore Foundation through Grant GBMF5076. F. Daigne, R. Mochkovitch and R. Duque acknowledge material and financial support from the Centre National d'\'Etudes Spatiales.



\appendix
\section{Estimating the maximal spreading of the jet during the plateau phase}
\label{sec:spread}

We denote by $\theta_{j,0}$ the initial opening angle of the core, and $\theta_{j,f}$ its value at the end of the plateau phase.
For the evolution of~$\theta_{j}$ we adopt the prescription of \cite{GP2012}, using their Eq.~13 with $a=1$, which provides a satisfactory fit of the numerical simulations, 
\begin{equation}
\label{eq:A1}
{d\theta_j\over dR} \approx {1\over R\,\Gamma^{2}\,\theta_j}\, .
\end{equation}
After the onset of deceleration of the core, we can write $\Gamma\propto R^{-\epsilon}$ with $\epsilon=3/2$ (resp. $1/2)$ in the uniform medium (resp. wind) case.
These scalings, which were derived for a spherical outflow, are justified only if lateral spreading is ultimately found to be negligible. As we show below this is the case for all cases of interest in this work.
Using these relations we integrate Eq.~\ref{eq:A1} to:
\begin{equation}
\label{eq:A2}
\theta_{j,f}^2 - \theta_{j,0}^2= \frac{1}{\epsilon}\left(\frac{1}{\Gamma_p^2}-\frac{1}{\Gamma_j^2}\right)
\end{equation}
where we have denoted by $\Gamma_p$ the core's bulk Lorentz factor at the end of the plateau.

``De-beamed core plateaus'': Here the plateau ends when $\Gamma_p^{-1} = \theta_v - \theta_{j,f}$, so that
\begin{equation}
\theta_{j,f}^2 - \theta_{j,0}^2 = {1\over \epsilon}\,\left[(\theta_v-\theta_{j,f})^2-{1\over \Gamma_j^2}\right]\approx {1\over \epsilon}\,(\theta_v-\theta_{j,f})^2 \,  .
\end{equation}

\begin{table}
\begin{center}
\begin{tabular}{l|ll}
\hline
\hline
$\theta_{\rm v}$ &  $\theta_{j,f}^{u}$  &   $\theta_{j,f}^{w}$ 
\\
\hline
0.11 &  0.1003  & 0.1008  \\
0.12 &  0.1012  & 0.1029  \\
0.13 &  0.1025  & 0.1057 \\
\hline
\end{tabular}
\end{center}
\caption{Values of the jet opening angle at the end of the plateau phase, $\theta_{j,f}$ for three values of the viewing angle, $\theta_{\rm v}$. The results are given for the ``de-beamed core" model and for a uniform (superscript `u') or a wind (superscript `w') external medium. An initial value $\theta_{j,0}=0.1$ rad has been adopted in all cases.}
\label{tab:A1}
\end{table}
The results obtained by solving this equation for $\theta_{j,0}=0.1$ rad, $\Gamma_j=300$ and three values of the viewing angle
are given in Tab.~\ref{tab:A1}.

``Late deceleration plateaus'': Here, the end of the plateau is when the LOS material starts to decelerate, at a radius $R_d^{LOS}$. Seeing as $R_d \propto(E/\Gamma_0^2)^{1/(3-s)}$ ($s$ = 0 in ISM and $s$ = 2 in wind), two situations can occur.

(i) If $2\beta - \alpha < 0$, then the deceleration radius of the core jet is larger than that of LOS material, and thus the emission of the plateau radiation occurs during coasting of the core. In this case, Eq.~\ref{eq:A1} leads to only a logarithmic increase in $\theta_{\rm j}^2(R)$. The plateau here therefore expands negligibly. This is the case for the values of $\alpha \approx 3\beta$ found to match the observed correlations in \S \ref{sec:losplateau}.

(ii) If $2\beta - \alpha > 0$, the core may indeed expand according to Eq.~\ref{eq:A2} during the plateau phase, where its final Lorentz factor is estimated as:
\begin{equation}
\label{eq:a4}
\Gamma_p = \Gamma_j \left(\frac{R_d^{LOS}}{R_d^{\theta_j}}\right)^{-\epsilon}.
\end{equation}
Assuming $\theta_v > \theta_{j,0}$, we obtain: 
\begin{equation}
\theta_{j,f}^2 - \theta_{j,0}^2 = \frac{1}{\epsilon \Gamma_j^2}\left[\left(\frac{\theta_{v}}{\theta_{j,0}}\right)^{\frac{2 \epsilon (2\beta - \alpha)}{3 - s}} - 1\right]
\end{equation}
Evaluating this for $\alpha = 4$, $\beta = 3$, $\Delta \theta = 0.03$, $\Gamma_j = 100$, and assuming a wind environment, one obtains a relative increase of $\theta_j$ of $\lesssim$ 1\% during the plateau phase.

We conclude that in both models the jet spreading remains very limited until the end of the plateau. 
In addition, the prescription for the lateral spreading taken from 
\citet{GP2012} corresponds to a top-hat jet. With a lateral structure as discussed in the present paper, the confinement by the material at higher latitude should even further limit the spreading of the core region.


\bsp	
\label{lastpage}

\begin{thebibliography}{}
\makeatletter
\relax
\def\mn@urlcharsother{\let\do\@makeother \do\$\do\&\do\#\do\^\do\_\do\%\do\~}
\def\mn@doi{\begingroup\mn@urlcharsother \@ifnextchar [ {\mn@doi@}
  {\mn@doi@[]}}
\def\mn@doi@[#1]#2{\def\@tempa{#1}\ifx\@tempa\@empty \href
  {http://dx.doi.org/#2} {doi:#2}\else \href {http://dx.doi.org/#2} {#1}\fi
  \endgroup}
\def\mn@eprint#1#2{\mn@eprint@#1:#2::\@nil}
\def\mn@eprint@arXiv#1{\href {http://arxiv.org/abs/#1} {{\tt arXiv:#1}}}
\def\mn@eprint@dblp#1{\href {http://dblp.uni-trier.de/rec/bibtex/#1.xml}
  {dblp:#1}}
\def\mn@eprint@#1:#2:#3:#4\@nil{\def\@tempa {#1}\def\@tempb {#2}\def\@tempc
  {#3}\ifx \@tempc \@empty \let \@tempc \@tempb \let \@tempb \@tempa \fi \ifx
  \@tempb \@empty \def\@tempb {arXiv}\fi \@ifundefined
  {mn@eprint@\@tempb}{\@tempb:\@tempc}{\expandafter \expandafter \csname
  mn@eprint@\@tempb\endcsname \expandafter{\@tempc}}}

\bibitem[\protect\citeauthoryear{{Beloborodov}}{{Beloborodov}}{2005}]{Beloborodov2005}
{Beloborodov} A.~M.,  2005, \mn@doi [\apj] {10.1086/430166}, \href
  {https://ui.adsabs.harvard.edu/abs/2005ApJ...627..346B} {627, 346}

\bibitem[\protect\citeauthoryear{{Beniamini} \& {Mochkovitch}}{{Beniamini} \&
  {Mochkovitch}}{2017}]{BM2017}
{Beniamini} P.,  {Mochkovitch} R.,  2017, \mn@doi [\aap]
  {10.1051/0004-6361/201730523}, \href
  {http://cdsads.u-strasbg.fr/abs/2017A$\%$26A...605A..60B} {605, A60}

\bibitem[\protect\citeauthoryear{{Beniamini} \& {Nakar}}{{Beniamini} \&
  {Nakar}}{2019}]{BN2019}
{Beniamini} P.,  {Nakar} E.,  2019, \mn@doi [\mnras] {10.1093/mnras/sty3110},
  \href {http://cdsads.u-strasbg.fr/abs/2019MNRAS.482.5430B} {482, 5430}

\bibitem[\protect\citeauthoryear{{Beniamini} \& {van der Horst}}{{Beniamini} \&
  {van der Horst}}{2017}]{BvdH2017}
{Beniamini} P.,  {van der Horst} A.~J.,  2017, \mn@doi [\mnras]
  {10.1093/mnras/stx2203}, \href
  {http://adsabs.harvard.edu/abs/2017MNRAS.472.3161B} {472, 3161}

\bibitem[\protect\citeauthoryear{{Beniamini}, {Nava}, {Duran}  \&
  {Piran}}{{Beniamini} et~al.}{2015}]{Beniamini2015}
{Beniamini} P.,  {Nava} L.,  {Duran} R.~B.,   {Piran} T.,  2015, \mn@doi
  [\mnras] {10.1093/mnras/stv2033}, \href
  {http://adsabs.harvard.edu/abs/2015MNRAS.454.1073B} {454, 1073}

\bibitem[\protect\citeauthoryear{{Beniamini}, {Nava}  \& {Piran}}{{Beniamini}
  et~al.}{2016}]{Beniamini2016}
{Beniamini} P.,  {Nava} L.,   {Piran} T.,  2016, \mn@doi [\mnras]
  {10.1093/mnras/stw1331}, \href
  {http://adsabs.harvard.edu/abs/2016MNRAS.461...51B} {461, 51}

\bibitem[\protect\citeauthoryear{{Beniamini}, {Petropoulou}, {Barniol Duran}
  \& {Giannios}}{{Beniamini} et~al.}{2019}]{Beniamini2019}
{Beniamini} P.,  {Petropoulou} M.,  {Barniol Duran} R.,   {Giannios} D.,  2019,
  \mn@doi [\mnras] {10.1093/mnras/sty3093}, \href
  {https://ui.adsabs.harvard.edu/abs/2019MNRAS.483..840B} {483, 840}

\bibitem[\protect\citeauthoryear{{Beniamini}, {Barniol Duran}, {Petropoulou}
  \& {Giannios}}{{Beniamini} et~al.}{2020}]{Beniamini2020}
{Beniamini} P.,  {Barniol Duran} R.,  {Petropoulou} M.,   {Giannios} D.,  2020,
  arXiv e-prints, \href {https://ui.adsabs.harvard.edu/abs/2020arXiv200100950B}
  {p. arXiv:2001.00950}

\bibitem[\protect\citeauthoryear{{Dainotti}, {Cardone}  \&
  {Capozziello}}{{Dainotti} et~al.}{2008}]{D08}
{Dainotti} M.~G.,  {Cardone} V.~F.,   {Capozziello} S.,  2008, \mn@doi [\mnras]
  {10.1111/j.1745-3933.2008.00560.x}, \href
  {https://ui.adsabs.harvard.edu/abs/2008MNRAS.391L..79D} {391, L79}

\bibitem[\protect\citeauthoryear{{Dainotti}, {Hernandez}, {Postnikov},
  {Nagataki}, {O'brien}, {Willingale}  \& {Striegel}}{{Dainotti}
  et~al.}{2017}]{D17}
{Dainotti} M.~G.,  {Hernandez} X.,  {Postnikov} S.,  {Nagataki} S.,  {O'brien}
  P.,  {Willingale} R.,   {Striegel} S.,  2017, \mn@doi [\apj]
  {10.3847/1538-4357/aa8a6b}, \href
  {https://ui.adsabs.harvard.edu/abs/2017ApJ...848...88D} {848, 88}

\bibitem[\protect\citeauthoryear{{Duffell} \& {MacFadyen}}{{Duffell} \&
  {MacFadyen}}{2015}]{Duffell2015}
{Duffell} P.~C.,  {MacFadyen} A.~I.,  2015, \mn@doi [\apj]
  {10.1088/0004-637X/806/2/205}, \href
  {https://ui.adsabs.harvard.edu/abs/2015ApJ...806..205D} {806, 205}

\bibitem[\protect\citeauthoryear{{Duque}, {Daigne}  \& {Mochkovitch}}{{Duque}
  et~al.}{2019}]{Duque2019}
{Duque} R.,  {Daigne} F.,   {Mochkovitch} R.,  2019, \mn@doi [\aap]
  {10.1051/0004-6361/201935926}, \href
  {https://ui.adsabs.harvard.edu/abs/2019A&A...631A..39D} {631, A39}

\bibitem[\protect\citeauthoryear{{Eichler} \& {Granot}}{{Eichler} \&
  {Granot}}{2006}]{Eichler2006}
{Eichler} D.,  {Granot} J.,  2006, \mn@doi [\apjl] {10.1086/503667}, \href
  {http://cdsads.u-strasbg.fr/abs/2006ApJ...641L...5E} {641, L5}

\bibitem[\protect\citeauthoryear{{Gehrels}}{{Gehrels}}{2004}]{Gehrels2004}
{Gehrels} N.,  2004, in {Fenimore} E.,  {Galassi} M.,  eds,  American Institute
  of Physics Conference Series Vol. 727, Gamma-Ray Bursts: 30 Years of
  Discovery. pp 637--641 (\mn@eprint {} {astro-ph/0405233}),
  \mn@doi{10.1063/1.1810924}

\bibitem[\protect\citeauthoryear{{Genet}, {Daigne}  \& {Mochkovitch}}{{Genet}
  et~al.}{2007}]{Genet2007}
{Genet} F.,  {Daigne} F.,   {Mochkovitch} R.,  2007, \mn@doi [\mnras]
  {10.1111/j.1365-2966.2007.12243.x}, \href
  {https://ui.adsabs.harvard.edu/abs/2007MNRAS.381..732G} {381, 732}

\bibitem[\protect\citeauthoryear{{Ghirlanda} et~al.,}{{Ghirlanda}
  et~al.}{2019}]{Ghirlanda2019}
{Ghirlanda} G.,  et~al., 2019, \mn@doi [Science] {10.1126/science.aau8815},
  \href {https://ui.adsabs.harvard.edu/abs/2019Sci...363..968G} {363, 968}

\bibitem[\protect\citeauthoryear{{Ghisellini}, {Ghirlanda}, {Nava}  \&
  {Firmani}}{{Ghisellini} et~al.}{2007}]{Ghisellini2007}
{Ghisellini} G.,  {Ghirlanda} G.,  {Nava} L.,   {Firmani} C.,  2007, \mn@doi
  [\apjl] {10.1086/515570}, \href
  {https://ui.adsabs.harvard.edu/abs/2007ApJ...658L..75G} {658, L75}

\bibitem[\protect\citeauthoryear{{Gill} \& {Granot}}{{Gill} \&
  {Granot}}{2018}]{Gill2018}
{Gill} R.,  {Granot} J.,  2018, \mn@doi [\mnras] {10.1093/mnras/sty1214}, \href
  {https://ui.adsabs.harvard.edu/abs/2018MNRAS.478.4128G} {478, 4128}

\bibitem[\protect\citeauthoryear{{Gill}, {Granot}, {De Colle}  \&
  {Urrutia}}{{Gill} et~al.}{2019}]{Gill2019}
{Gill} R.,  {Granot} J.,  {De Colle} F.,   {Urrutia} G.,  2019, \mn@doi [\apj]
  {10.3847/1538-4357/ab3577}, \href
  {https://ui.adsabs.harvard.edu/abs/2019ApJ...883...15G} {883, 15}

\bibitem[\protect\citeauthoryear{{Gottlieb}, {Levinson}  \& {Nakar}}{{Gottlieb}
  et~al.}{2019a}]{Gottlieb2019b}
{Gottlieb} O.,  {Levinson} A.,   {Nakar} E.,  2019a, \mn@doi [\mnras]
  {10.1093/mnras/stz1828}, \href
  {https://ui.adsabs.harvard.edu/abs/2019MNRAS.488.1416G} {488, 1416}

\bibitem[\protect\citeauthoryear{{Gottlieb}, {Nakar}  \& {Piran}}{{Gottlieb}
  et~al.}{2019b}]{Gottlieb2019}
{Gottlieb} O.,  {Nakar} E.,   {Piran} T.,  2019b, \mn@doi [\mnras]
  {10.1093/mnras/stz1906}, \href
  {https://ui.adsabs.harvard.edu/abs/2019MNRAS.488.2405G} {488, 2405}

\bibitem[\protect\citeauthoryear{{Granot} \& {Kumar}}{{Granot} \&
  {Kumar}}{2006}]{GK2006}
{Granot} J.,  {Kumar} P.,  2006, \mn@doi [\mnras]
  {10.1111/j.1745-3933.2005.00121.x}, \href
  {https://ui.adsabs.harvard.edu/abs/2006MNRAS.366L..13G} {366, L13}

\bibitem[\protect\citeauthoryear{{Granot} \& {Piran}}{{Granot} \&
  {Piran}}{2012}]{GP2012}
{Granot} J.,  {Piran} T.,  2012, \mn@doi [\mnras]
  {10.1111/j.1365-2966.2011.20335.x}, \href
  {https://ui.adsabs.harvard.edu/abs/2012MNRAS.421..570G} {421, 570}

\bibitem[\protect\citeauthoryear{{Granot} \& {Sari}}{{Granot} \&
  {Sari}}{2002}]{GS2002}
{Granot} J.,  {Sari} R.,  2002, \mn@doi [\apj] {10.1086/338966}, \href
  {http://adsabs.harvard.edu/abs/2002ApJ...568..820G} {568, 820}

\bibitem[\protect\citeauthoryear{{Granot} \& {van der Horst}}{{Granot} \& {van
  der Horst}}{2014}]{GvdH2014}
{Granot} J.,  {van der Horst} A.~J.,  2014, \mn@doi [\pasa]
  {10.1017/pasa.2013.44}, \href
  {http://adsabs.harvard.edu/abs/2014PASA...31....8G} {31, e008}

\bibitem[\protect\citeauthoryear{{Granot}, {K{\"o}nigl}  \& {Piran}}{{Granot}
  et~al.}{2006}]{Granot2006}
{Granot} J.,  {K{\"o}nigl} A.,   {Piran} T.,  2006, \mn@doi [\mnras]
  {10.1111/j.1365-2966.2006.10621.x}, \href
  {https://ui.adsabs.harvard.edu/abs/2006MNRAS.370.1946G} {370, 1946}

\bibitem[\protect\citeauthoryear{{Granot}, {Gill}, {Guetta}  \& {De
  Colle}}{{Granot} et~al.}{2018}]{Granot2018}
{Granot} J.,  {Gill} R.,  {Guetta} D.,   {De Colle} F.,  2018, \mn@doi [\mnras]
  {10.1093/mnras/sty2308}, \href
  {http://cdsads.u-strasbg.fr/abs/2018MNRAS.481.1597G} {481, 1597}

\bibitem[\protect\citeauthoryear{{Hasco{\"e}t}, {Daigne}  \&
  {Mochkovitch}}{{Hasco{\"e}t} et~al.}{2014}]{Hascoet2014}
{Hasco{\"e}t} R.,  {Daigne} F.,   {Mochkovitch} R.,  2014, \mn@doi [\mnras]
  {10.1093/mnras/stu750}, \href
  {https://ui.adsabs.harvard.edu/abs/2014MNRAS.442...20H} {442, 20}

\bibitem[\protect\citeauthoryear{{Ioka} \& {Nakamura}}{{Ioka} \&
  {Nakamura}}{2018}]{Ioka2018}
{Ioka} K.,  {Nakamura} T.,  2018, \mn@doi [Progress of Theoretical and
  Experimental Physics] {10.1093/ptep/pty036}, \href
  {http://adsabs.harvard.edu/abs/2018PTEP.2018d3E02I} {2018, 043E02}

\bibitem[\protect\citeauthoryear{{Ioka}, {Toma}, {Yamazaki}  \&
  {Nakamura}}{{Ioka} et~al.}{2006}]{Ioka2006}
{Ioka} K.,  {Toma} K.,  {Yamazaki} R.,   {Nakamura} T.,  2006, \mn@doi [\aap]
  {10.1051/0004-6361:20064939}, \href
  {https://ui.adsabs.harvard.edu/abs/2006A%26A...458....7I} {458, 7}

\bibitem[\protect\citeauthoryear{{Kasliwal} et~al.,}{{Kasliwal}
  et~al.}{2017}]{Kasliwal2017}
{Kasliwal} M.~M.,  et~al., 2017, \mn@doi [Science] {10.1126/science.aap9455},
  \href {http://adsabs.harvard.edu/abs/2017Sci...358.1559K} {358, 1559}

\bibitem[\protect\citeauthoryear{{Kathirgamaraju}, {Barniol Duran}  \&
  {Giannios}}{{Kathirgamaraju} et~al.}{2018}]{Kathirgamaraju2018}
{Kathirgamaraju} A.,  {Barniol Duran} R.,   {Giannios} D.,  2018, \mn@doi
  [\mnras] {10.1093/mnrasl/slx175}, \href
  {http://adsabs.harvard.edu/abs/2018MNRAS.473L.121K} {473, L121}

\bibitem[\protect\citeauthoryear{{Kobayashi} \& {Zhang}}{{Kobayashi} \&
  {Zhang}}{2007}]{Kobayashi2007}
{Kobayashi} S.,  {Zhang} B.,  2007, \mn@doi [\apj] {10.1086/510203}, \href
  {https://ui.adsabs.harvard.edu/abs/2007ApJ...655..973K} {655, 973}

\bibitem[\protect\citeauthoryear{{Kumar} \& {Panaitescu}}{{Kumar} \&
  {Panaitescu}}{2000}]{KP2000}
{Kumar} P.,  {Panaitescu} A.,  2000, \mn@doi [\apjl] {10.1086/312905}, \href
  {https://ui.adsabs.harvard.edu/abs/2000ApJ...541L..51K} {541, L51}

\bibitem[\protect\citeauthoryear{{Kumar} \& {Zhang}}{{Kumar} \&
  {Zhang}}{2015}]{Kumar2015}
{Kumar} P.,  {Zhang} B.,  2015, \mn@doi [\physrep]
  {10.1016/j.physrep.2014.09.008}, \href
  {http://cdsads.u-strasbg.fr/abs/2015PhR...561....1K} {561, 1}

\bibitem[\protect\citeauthoryear{{Lamb} \& {Kobayashi}}{{Lamb} \&
  {Kobayashi}}{2017}]{Lamb2017}
{Lamb} G.~P.,  {Kobayashi} S.,  2017, \mn@doi [\mnras] {10.1093/mnras/stx2345},
  \href {http://adsabs.harvard.edu/abs/2017MNRAS.472.4953L} {472, 4953}

\bibitem[\protect\citeauthoryear{{Leventis}, {Wijers}  \& {van der
  Horst}}{{Leventis} et~al.}{2014}]{Leventis2014}
{Leventis} K.,  {Wijers} R.~A.~M.~J.,   {van der Horst} A.~J.,  2014, \mn@doi
  [\mnras] {10.1093/mnras/stt2055}, \href
  {https://ui.adsabs.harvard.edu/abs/2014MNRAS.437.2448L} {437, 2448}

\bibitem[\protect\citeauthoryear{{Li} et~al.,}{{Li} et~al.}{2012}]{Li2012}
{Li} L.,  et~al., 2012, \mn@doi [\apj] {10.1088/0004-637X/758/1/27}, \href
  {https://ui.adsabs.harvard.edu/abs/2012ApJ...758...27L} {758, 27}

\bibitem[\protect\citeauthoryear{{Liang} et~al.,}{{Liang}
  et~al.}{2006}]{Liang2006}
{Liang} E.~W.,  et~al., 2006, \mn@doi [\apj] {10.1086/504684}, \href
  {http://cdsads.u-strasbg.fr/abs/2006ApJ...646..351L} {646, 351}

\bibitem[\protect\citeauthoryear{{Liang} et~al.,}{{Liang}
  et~al.}{2013}]{Liang2013}
{Liang} E.-W.,  et~al., 2013, \mn@doi [\apj] {10.1088/0004-637X/774/1/13},
  \href {https://ui.adsabs.harvard.edu/abs/2013ApJ...774...13L} {774, 13}

\bibitem[\protect\citeauthoryear{{Margutti} et~al.,}{{Margutti}
  et~al.}{2013}]{Margutti2013}
{Margutti} R.,  et~al., 2013, \mn@doi [\mnras] {10.1093/mnras/sts066}, \href
  {http://adsabs.harvard.edu/abs/2013MNRAS.428..729M} {428, 729}

\bibitem[\protect\citeauthoryear{{Nava} et~al.,}{{Nava}
  et~al.}{2014}]{Nava2014}
{Nava} L.,  et~al., 2014, \mn@doi [\mnras] {10.1093/mnras/stu1451}, \href
  {http://adsabs.harvard.edu/abs/2014MNRAS.443.3578N} {443, 3578}

\bibitem[\protect\citeauthoryear{{Nousek} et~al.,}{{Nousek}
  et~al.}{2006}]{Nousek2006}
{Nousek} J.~A.,  et~al., 2006, \mn@doi [\apj] {10.1086/500724}, \href
  {http://cdsads.u-strasbg.fr/abs/2006ApJ...642..389N} {642, 389}

\bibitem[\protect\citeauthoryear{{Oganesyan}, {Ascenzi}, {Branchesi}, {Sharan
  Salafia}, {Dall'Osso}  \& {Ghirlanda}}{{Oganesyan}
  et~al.}{2019}]{Oganesyan2019}
{Oganesyan} G.,  {Ascenzi} S.,  {Branchesi} M.,  {Sharan Salafia} O.,
  {Dall'Osso} S.,   {Ghirlanda} G.,  2019, arXiv e-prints, \href
  {https://ui.adsabs.harvard.edu/abs/2019arXiv190408786O} {p. arXiv:1904.08786}

\bibitem[\protect\citeauthoryear{{Panaitescu}, {M{\'e}sz{\'a}ros}, {Burrows},
  {Nousek}, {Gehrels}, {O'Brien}  \& {Willingale}}{{Panaitescu}
  et~al.}{2006}]{Panaitescu2006}
{Panaitescu} A.,  {M{\'e}sz{\'a}ros} P.,  {Burrows} D.,  {Nousek} J.,
  {Gehrels} N.,  {O'Brien} P.,   {Willingale} R.,  2006, \mn@doi [\mnras]
  {10.1111/j.1365-2966.2006.10453.x}, \href
  {https://ui.adsabs.harvard.edu/abs/2006MNRAS.369.2059P} {369, 2059}

\bibitem[\protect\citeauthoryear{{Santana}, {Barniol Duran}  \&
  {Kumar}}{{Santana} et~al.}{2014}]{Santana2014}
{Santana} R.,  {Barniol Duran} R.,   {Kumar} P.,  2014, \mn@doi [\apj]
  {10.1088/0004-637X/785/1/29}, \href
  {http://adsabs.harvard.edu/abs/2014ApJ...785...29S} {785, 29}

\bibitem[\protect\citeauthoryear{{Sari} \& {Piran}}{{Sari} \&
  {Piran}}{1995}]{SariPiran1995}
{Sari} R.,  {Piran} T.,  1995, \mn@doi [\apjl] {10.1086/309835}, \href
  {https://ui.adsabs.harvard.edu/abs/1995ApJ...455L.143S} {455, L143}

\bibitem[\protect\citeauthoryear{{Sari}, {Piran}  \& {Narayan}}{{Sari}
  et~al.}{1998}]{Sari1998}
{Sari} R.,  {Piran} T.,   {Narayan} R.,  1998, \mn@doi [\apjl]
  {10.1086/311269}, \href
  {https://ui.adsabs.harvard.edu/abs/1998ApJ...497L..17S} {497, L17}

\bibitem[\protect\citeauthoryear{{Shen} \& {Matzner}}{{Shen} \&
  {Matzner}}{2012}]{Shen2012}
{Shen} R.,  {Matzner} C.~D.,  2012, \mn@doi [\apj]
  {10.1088/0004-637X/744/1/36}, \href
  {https://ui.adsabs.harvard.edu/abs/2012ApJ...744...36S} {744, 36}

\bibitem[\protect\citeauthoryear{{Tang}, {Huang}, {Geng}  \& {Zhang}}{{Tang}
  et~al.}{2019}]{Tang2019}
{Tang} C.-H.,  {Huang} Y.-F.,  {Geng} J.-J.,   {Zhang} Z.-B.,  2019, \mn@doi
  [\apjs] {10.3847/1538-4365/ab4711}, \href
  {https://ui.adsabs.harvard.edu/abs/2019ApJS..245....1T} {245, 1}

\bibitem[\protect\citeauthoryear{{Toma}, {Ioka}, {Yamazaki}  \&
  {Nakamura}}{{Toma} et~al.}{2006}]{Toma2006}
{Toma} K.,  {Ioka} K.,  {Yamazaki} R.,   {Nakamura} T.,  2006, \mn@doi [\apjl]
  {10.1086/503384}, \href
  {https://ui.adsabs.harvard.edu/abs/2006ApJ...640L.139T} {640, L139}

\bibitem[\protect\citeauthoryear{{Troja} et~al.,}{{Troja}
  et~al.}{2007}]{Troja2007}
{Troja} E.,  et~al., 2007, \mn@doi [\apj] {10.1086/519450}, \href
  {http://cdsads.u-strasbg.fr/abs/2007ApJ...665..599T} {665, 599}

\bibitem[\protect\citeauthoryear{{Uhm} \& {Beloborodov}}{{Uhm} \&
  {Beloborodov}}{2007}]{Uhm2007}
{Uhm} Z.~L.,  {Beloborodov} A.~M.,  2007, \mn@doi [\apj] {10.1086/519837},
  \href {https://ui.adsabs.harvard.edu/abs/2007ApJ...665L..93U} {665, L93}

\bibitem[\protect\citeauthoryear{{Uhm}, {Zhang}, {Hasco{\"e}t}, {Daigne},
  {Mochkovitch}  \& {Park}}{{Uhm} et~al.}{2012}]{Uhm2012}
{Uhm} Z.~L.,  {Zhang} B.,  {Hasco{\"e}t} R.,  {Daigne} F.,  {Mochkovitch} R.,
  {Park} I.~H.,  2012, \mn@doi [\apj] {10.1088/0004-637X/761/2/147}, \href
  {https://ui.adsabs.harvard.edu/abs/2012ApJ...761..147U} {761, 147}

\bibitem[\protect\citeauthoryear{{Wijers} \& {Galama}}{{Wijers} \&
  {Galama}}{1999}]{WG1999}
{Wijers} R.~A.~M.~J.,  {Galama} T.~J.,  1999, \mn@doi [\apj] {10.1086/307705},
  \href {https://ui.adsabs.harvard.edu/abs/1999ApJ...523..177W} {523, 177}

\bibitem[\protect\citeauthoryear{{Zhang}, {Fan}, {Dyks}, {Kobayashi},
  {M{\'e}sz{\'a}ros}, {Burrows}, {Nousek}  \& {Gehrels}}{{Zhang}
  et~al.}{2006}]{Zhang2006}
{Zhang} B.,  {Fan} Y.~Z.,  {Dyks} J.,  {Kobayashi} S.,  {M{\'e}sz{\'a}ros} P.,
  {Burrows} D.~N.,  {Nousek} J.~A.,   {Gehrels} N.,  2006, \mn@doi [\apj]
  {10.1086/500723}, \href {http://cdsads.u-strasbg.fr/abs/2006ApJ...642..354Z}
  {642, 354}

\bibitem[\protect\citeauthoryear{{Zhang}, {van Eerten}, {Burrows}, {Ryan},
  {Evans}, {Racusin}, {Troja}  \& {MacFadyen}}{{Zhang}
  et~al.}{2015}]{Zhang2015}
{Zhang} B.-B.,  {van Eerten} H.,  {Burrows} D.~N.,  {Ryan} G.~S.,  {Evans}
  P.~A.,  {Racusin} J.~L.,  {Troja} E.,   {MacFadyen} A.,  2015, \mn@doi [\apj]
  {10.1088/0004-637X/806/1/15}, \href
  {http://adsabs.harvard.edu/abs/2015ApJ...806...15Z} {806, 15}

\makeatother
\end{thebibliography}
\end{document}